\documentclass[
 reprint,
 amsmath, amssymb,
 aps,
prb,
]{revtex4-2}

\usepackage{graphicx}
\usepackage{dcolumn}
\usepackage{bm}
\usepackage{multirow}
\usepackage{braket}
\usepackage{color}
\usepackage{ulem}
\usepackage{txfonts}
\usepackage{longtable}
\usepackage{booktabs}
\usepackage{multirow}
\usepackage{comment}

\allowdisplaybreaks[4]

\begin{document}

\title{Symmetry-adapted closest Wannier modeling based on complete multipole basis set}

\author{Rikuto Oiwa$^1$, Akane Inda$^2$, Satoru Hayami$^2$, Takuya Nomoto$^3$, Ryotaro Arita$^{1,4}$, and Hiroaki Kusunose$^{5}$}
\affiliation{
$^1$RIKEN Center for Emergent Matter Science, Wako, Saitama 351-0198, Japan \\
$^2$Graduate School of Science, Hokkaido University, Sapporo 060-0810, Japan \\
$^3$Department of Physics, Tokyo Metropolitan University, Hachioji, Tokyo 192-0397, Japan \\
$^4$Department of Physics, University of Tokyo, Hongo, Tokyo 113-0033, Japan \\
$^5$Department of Physics, Meiji University, Kawasaki 214-8571, Japan
}

\begin{abstract}
We have developed a method to construct a symmetry-adapted Wannier tight-binding model based on the closest Wannier formalism and the symmetry-adapted multipole theory.
Since the symmetry properties of the closest Wannier functions are common to those of the original atomic orbitals, symmetry-adapted multipole basis (SAMB) can be defined as the complete orthonormal matrix basis set in the Hilbert space of the closest Wannier functions.
Utilizing the completeness and orthonormality of SAMBs, the closest Wannier Hamiltonian can be expressed as a linear combination of SAMBs belonging to the identity irreducible representation, thereby fully restoring the symmetry of the system.
Moreover, the linear coefficients of each SAMB (model parameters) related to crystalline electric fields, spin-orbit coupling, and electron hoppings are determined through simple matrix projection without any iterative procedure.
Thus, this method allows us to unveil mutual interplay among hidden electronic multipole degrees of freedom in the Hamiltonian and numerically evaluate them.
We demonstrate the effectiveness of our method by modeling monolayer graphene under a perpendicular electric field, highlighting its utility in symmetrizing the closest Wannier model, quantifying symmetry breaking, and predicting unusual responses.
The method is implemented in the open-source Python library SymClosestWannier, with the codes available on GitHub (https://github.com/CMT-MU/SymClosestWannier).
\end{abstract}

\maketitle

\section{Introduction}
\label{sec_intro}

A variety of material properties arises from the mutual interplay among electronic degrees of freedom; charge, atomic orbital and spin, and sublattices in molecules or crystals.
In particular, unconventional phase transitions and responses driven by hidden order parameters of charge-spin-orbital-sublattice composites have been extensively studied, e.g., hidden order in URu$_{2}$Si$_{2}$~\cite{fujimotoSpinNematicState2011, ikedaEmergentRank5Nematic2012c, chandraHastaticOrderHeavyfermion2013, PhysRevB.97.235142, Kambe_SCES_2019, Hayami_URu2Si2_2023}, ferro-axial orderings~\cite{Jin_RbFeMoO42_2020, Hayashida_RbFeMoO42_2021, Hayashida_ferro_rotation_2021, Hasegawa_ferro_rotation_2020, Hanate_ferro_rotation_2021, SH_ferro_rotation_2023, PhysRevLett.108.067201, PhysRevB.105.184407}, large anomalous Hall/Nernst effect under the cluster magnetic octupole ordering in Mn$_{3}$Sn~\cite{PhysRevB.95.094406, nakatsujiLargeAnomalousHall2015a, ikhlasLargeAnomalousNernst2017, kurodaEvidenceMagneticWeyl2017b}, and so on.

To accurately analyze material properties and predict new phenomena, constructing a reliable microscopic model is essential.
For this purpose, the tight-binding (TB) modeling based on the Wannier functions (WFs)~\cite{PhysRevB.56.12847, PhysRevB.65.035109,RevModPhys.84.1419,MOSTOFI2008685,MOSTOFI20142309,Pizzi_2020} derived from density-functional theory (DFT)~\cite{Hohenberg_Kohn_1964, Kohn_Sham_1965} is widely used, and it has become the {\it de facto} standard in modern materials science.
In particular, the Maximally localized WFs approach is the most popular technique.
Once a Wannier TB model is obtained, one can analyze quantitatively the atomic orbital and spin distributions in the electronic band structure.
By examining the symmetry and magnitude of the hopping integrals between WFs, one can infer the bonding feature, such as the directionality and strength of covalent-bonds.
Furthermore, the Wannier TB model can be utilized to study phase transitions, such as superconductivity and magnetism, as well as various external responses and many-body effects, including electron correlations and electron-phonon interactions.

Although a conventional Wannier TB modeling has proven effective in many applications, it faces three challenges, as outlined below:
(1) The conventional Wannier Hamiltonian often fails to rigorously respect the symmetry of the system, leading to slight lifting of band degeneracies on the scale of meV to $\mu$eV~\cite{PhysRevB.87.235109, KORETSUNE2023108645, PhysRevMaterials.2.103805}.
This issue primarily arises from the disentangling process within a specified energy window, the wannierization procedure, as well as numerical errors inherent in the DFT calculations.
This seemingly trivial issue can result in significant misinterpretations, especially in the analysis of band structures and phenomena that are sensitive to symmetry, band topologies, Berry curvature physics, and so on.
For instance, near the K point of graphene, relativistic spin-orbit coupling induces a tiny band gap on the order of $\mu$eV~\cite{Min_PRB_2006, Yao_PRB_2007}, while an electric field perpendicular to the plane introduces additional Rashba spin splitting~\cite{Gmitra_PRB_2009}, giving rise to the Rashba-Edelstein effect~\cite{PhysRevB.89.075422, PhysRevLett.119.196801,ghiasiChargetoSpinConversionRashba2019,PhysRevB.104.235429}.
Consequently, unexpected band splittings on the order of $\mu$eV lead to critical misinterpretations in both qualitative and quantitative analyses of the band structure and responses.
(2) The symmetry properties of WFs generally differ from those of the original atomic orbitals.
Consequently, the representation matrices for physical quantities, such as the orbital angular momentum operator, become ambiguous.
(3) The conventional Wannier TB modeling still faces challenges in clarifying the microscopic origins of orderings and responses.
Various unconventional ordering and responses may arise from hidden coupling among charge, atomic orbital and spin, and sublattice degrees of freedom. 
While a model Hamiltonian can be numerically obtained using conventional Wannier TB modeling, extracting these hidden couplings remains unclear.
Such microscopic information is crucial in understanding the microscopic origins of orderings and responses, and exploring further new functional materials.

R. Sakuma has proposed a method for constructing symmetry-adapted WFs~\cite{PhysRevB.87.235109}, by incorporating symmetry constraints into the wannierization process.
The Wannier Hamiltonian is then symmetrized by using the obtained symmetry-adapted WFs.
However, this method is not applicable when a frozen energy window is specified during the disentangling procedure.
Recently, T. Koretsune proposed another algorithm to create symmetry-adapted WFs which is applicable to the frozen window technique~\cite{KORETSUNE2023108645}.
This approach enables us to construct symmetry-adapted Wannier model that reproduces DFT band structure inside a frozen window.

On the other hand, D. Gresch {\it et al.} have proposed another approach to directly symmetrize the Wannier Hamiltonian by utilizing the symmetry properties of the WFs~\cite{PhysRevMaterials.2.103805, ZHI2022108196}.
In this method, the WFs need to have common symmetry properties as the original atomic orbitals~\cite{ZHI2022108196}.
However, this is usually difficult to be realized as mentioned in the problem (2), since the symmetry property of the WFs becomes unclear after wannierization process.

Meanwhile, T. Ozaki has recently developed a non-iterative approach to construct the closest Wannier (CW) functions (CWFs) to a given set of the original atomic orbitals~\cite{Ozaki_CW_2024}.
In this formalism, the disentanglement of bands is achieved with non-iterative calculations, significantly reducing computational costs, and the symmetry properties of the CWFs become common to those of the original atomic orbitals.
As a result, the CW Hamiltonian satisfies the symmetry of the system except for inherent numerical errors caused by the DFT calculations and the wannierization process.
From the perspectives of symmetry and computational cost, the CW model provides a good starting point to overcome the above three problems.

\begin{table}[t]
  \renewcommand{\arraystretch}{1.2}
  \begin{center}
  \caption{ \label{tab_samb_hamiltonian}
  Correspondence between the physical quantities appeared in Hamiltonian and the symmetry-adapted multipole basis (SAMB)~\cite{PhysRevB.107.195118}.
  $\mathbb{X}_{lm}^{\rm (a)}$ ($\mathbb{X} = \mathbb{Q}, \mathbb{M}, \mathbb{T}, \mathbb{G}$) and $\mathbb{Y}_{lm}^{\rm (s/b)}$ ($\mathbb{Y} = \mathbb{Q}, \mathbb{T}$) represent the atomic and site/bond-cluster SAMBs, respectively.
  The upper and lower panels represent on-site and hopping terms, and the site dependence in the upper panel is expressed by $\mathbb{Q}_{lm}^{\rm (s)}$.
  The detailed meaning of these symbols will be explained in Sec.~\ref{sec_symcw}.
  In the lines of Rashba and chiral SOC, $b^{\mu}_{ij}$ denotes the $\mu$ component of the bond vector from $j$ site to $i$ site.
  Note that the repeated indices are implicitly summed in the expression.
  }
  \begin{ruledtabular}
  \begin{tabular}{lcc}
    Type & Expression & Correspondence \\
    \hline
    Electric potential & $\phi q$ & $\phi \mathbb{Q}_{0}^{\rm (a)}$
    \\
    Crystal field & $\phi_{l m} Q_{l m}$ & $\phi_{l m} \mathbb{Q}_{l m}^{\rm (a)}$
    \\
    Zeeman term & $-h^{\mu} m^{\mu}$ & $-h^{\mu} \mathbb{M}_{1 m}^{\rm (a)}$
    \\
    Atomic SOC & $\zeta l^{\mu} \sigma^{\mu}$ & $\zeta \mathbb{Q}_{0, (1,1)}^{\rm (a)}$
    \\
    \hline
    Real hopping & $t_{i j} (c_i^{\dagger} c_j+$ H.c.) & $t_{i j} \mathbb{Q}_{l m}^{(\mathrm{b})}$
    \\
    Imaginary hopping & $t_{i j} (\mathrm{i} c_i^{\dagger} c_j+$ H.c.) & $t_{i j} \mathbb{T}_{l m}^{(\mathrm{b})}$
    \\
    Rashba SOC & $ \lambda_{\rm R} \epsilon_{z\mu\nu} (\mathrm{i} c_{i,\sigma}^{\dagger} b_{ij}^{\mu} \sigma_{\sigma\sigma'}^{\nu} c_{j,\sigma'} + {\rm H.c.})$ & $\lambda_{\rm R} \epsilon_{z\mu\nu} \mathbb{M}_{\mu}^{\rm (a)} \mathbb{T}_{\nu}^{\rm (b)}$
      \\
    Chiral SOC & $ \lambda_{\rm C} (\mathrm{i} c_{i,\sigma}^{\dagger} b_{ij}^{\mu} \sigma_{\sigma\sigma'}^{\mu} c_{j,\sigma'} + {\rm H.c.})$ & $\lambda_{\rm C} \mathbb{M}_{\mu}^{\rm (a)} \mathbb{T}_{\mu}^{\rm (b)}$
  \end{tabular}
  \end{ruledtabular}
  \end{center}
\end{table}

\begin{table}[t]
  \renewcommand{\arraystretch}{1.2}
  \begin{center}
  \caption{ \label{tab_samb_external_field_response}
  The spatial-inversion and time-reversal parities of (upper panel) the external fields and (lower panel) the responses and their correspondence to the SAMBs~\cite{PhysRevB.104.054412, SH_JPSJ_review_2024}.
  In the column of multipole, $\mathbb{X}_{lm}$ $(l = 0-3)$ means the rank-$l$ multipole ($\mathbb{X} = \mathbb{Q},\mathbb{M},\mathbb{T},\mathbb{G})$.
  }
  \begin{ruledtabular}
  \begin{tabular}{lcccc}
  Max. rank & $\mathcal{P}$ & $\mathcal{T}$ & External field & Multipole
  \\
  \hline
  0 & $+$ & $+$ & $\bm{\nabla} \cdot \bm{E}$ & $\mathbb{Q}_0$
  \\
    & $-$ & $-$ & $\bm{E} \cdot \bm{B}$ & $\mathbb{M}_0$
  \\
    & $+$ & $-$ & $\bm{B} \cdot(\bm{\nabla} \times \bm{E}), \bm{E} \cdot(\bm{\nabla} \times \bm{B})$ & $\mathbb{T}_0$
  \\
    & $-$ & $+$ & $\bm{E} \cdot(\bm{\nabla} \times \bm{E}), \bm{B} \cdot(\bm{\nabla} \times \bm{B})$ & $\mathbb{G}_0$
  \\
  \hline
  1 & $-$ & $+$ & electric field $\bm{E}$ & $\mathbb{Q}_{1m}$
  \\
    &     &     & thermal gradient $-\bm{\nabla} T$ & $\mathbb{Q}_{1m}$
  \\
    & $+$ & $-$ & magnetic field $\bm{H}$ & $\mathbb{M}_{1m}$
  \\
    & $-$ & $-$ & $\bm{\nabla} \times \bm{B}, \frac{\partial \bm{E}}{\partial t}$ & $\mathbb{T}_{1m}$
    \\
    & $+$ & $+$ & $\bm{\nabla} \times \bm{E}, \frac{\partial \bm{B}}{\partial t}$ & $\mathbb{G}_{1m}$
    \\
  \hline
  2 & $+$ & $+$ & stress $\tau_{i j}$ & $\mathbb{Q}_0, \mathbb{Q}_{2m}$
  \\
    & $+$ & $+$ & nonlinear field $E_i E_j\left(H_i H_j\right)$ & $\mathbb{Q}_0, \mathbb{Q}_{2m}$
    \\
    & $-$ & $-$ & composite field $E_i H_j$ & $\mathbb{M}_0, \mathbb{T}_{1m}, \mathbb{M}_{2m}$
    \\
  \hline
  3 & $-$ & $+$ & nonlinear field $E_i E_j E_k$ & $\mathbb{Q}_{1m}, \mathbb{Q}_{3m}$
    \\
    & $+$ & $-$ & nonlinear field $H_i H_j H_k$ & $\mathbb{M}_{1m}, \mathbb{M}_{3m}$
    \\
    \hline
Max. rank & $\mathcal{P}$ & $\mathcal{T}$ & Response & Multipole
    \\
    \hline
  0 & $+$ & $+$ & temperature change $\Delta T$ & $\mathbb{Q}_0$
    \\
    & $-$ & $+$ & chirality & $\mathbb{G}_0$
    \\
    \hline
  1 & $-$ & $+$ & electric polarization $\bm{P}$ & $\mathbb{Q}_{1m}$
    \\
    & $+$ & $-$ & magnetization $\bm{M}$ & $\mathbb{M}_{1m}$
    \\
    & $-$ & $-$ & electric (thermal) current $\bm{J}\left(\bm{J}^{\mathrm{Q}}\right)$ & $\mathbb{T}_{1m}$
    \\
    & $+$ & $+$ & rotational distortion $\bm{\omega}$ & $\mathbb{G}_{1m}$
    \\
    \hline
  2 & $+$ & $+$ & strain $\bm{\varepsilon}$ & $\mathbb{Q}_0, \mathbb{Q}_{2m}$
    \\
    & $-$ & $+$ & spin current $J_i \sigma_j$ & $\mathbb{G}_0, \mathbb{Q}_{1m}, \mathbb{G}_{2m}$
\end{tabular}
\end{ruledtabular}
\end{center}
\end{table}

In the present study, we propose a post-processing symmetrization step after the construction of the CW model based on the symmetry-adapted multipole basis (SAMB)~\cite{JPSJ.87.033709, PhysRevB.98.165110, PhysRevB.102.144441, JPSJ.89.104704}.
According to the spatial-inversion and time-reversal parities, SAMBs are classified into four types of multipoles: electric (E), magnetic (M), magnetic toroidal (MT), and electric toroidal (ET) multipoles~\cite{JPSJ.87.033709, PhysRevB.98.165110, JPSJ.89.104704}.
Each SAMB is further characterized by the rank of multipole, the irreducible representation and its component under the point group symmetry of the system.
Since a set of SAMBs is complete and orthonormal in the given Hilbert space of molecules or crystals, it describes any parity-specific anisotropic distributions of composite electronic and lattice degrees of freedom~\cite{PhysRevB.107.195118}.
Considering the symmetry properties of the CWFs are common to that of the original atomic orbitals, SAMBs can be defined as the complete orthonormal matrix basis set in the Hilbert space of the CWFs.
Then by utilizing the completeness and orthonormality of SAMBs, the CW Hamiltonian is expressed as the linear combination of SAMBs belonging to the identity irreducible representation, which are compatible to the symmetry of the system.
In contrast to the previous methods~\cite{PhysRevB.107.195118, RO_PRL_2022}, the linear coefficients of each SAMB, corresponding to the model parameters, are determined by a simple matrix projection without any iterative process, significantly decreasing computational costs.

As summarized in Table~\ref{tab_samb_hamiltonian}, each SAMB has the clear physical meaning, e.g., the crystalline electric field (CEF), spin-orbit coupling (SOC), or electron hoppings~\cite{PhysRevB.107.195118}, and so on, in symmetry-adapted manner.
Moreover, our method gives us insight into various unconventional responses driven by electronic multipole degrees of freedom hidden in the CW Hamiltonian, because there is a one-to-one correspondence between SAMBs and various external fields and responses~\cite{PhysRevB.104.054412, SH_JPSJ_review_2024}, examples of which are summarized in Table~\ref{tab_samb_external_field_response}.
The organization of the paper is given in the next outline section.

\section{Outline}
\label{sec_outline}

\begin{figure}[t]
  \begin{center}
  \includegraphics[width=0.98\hsize]{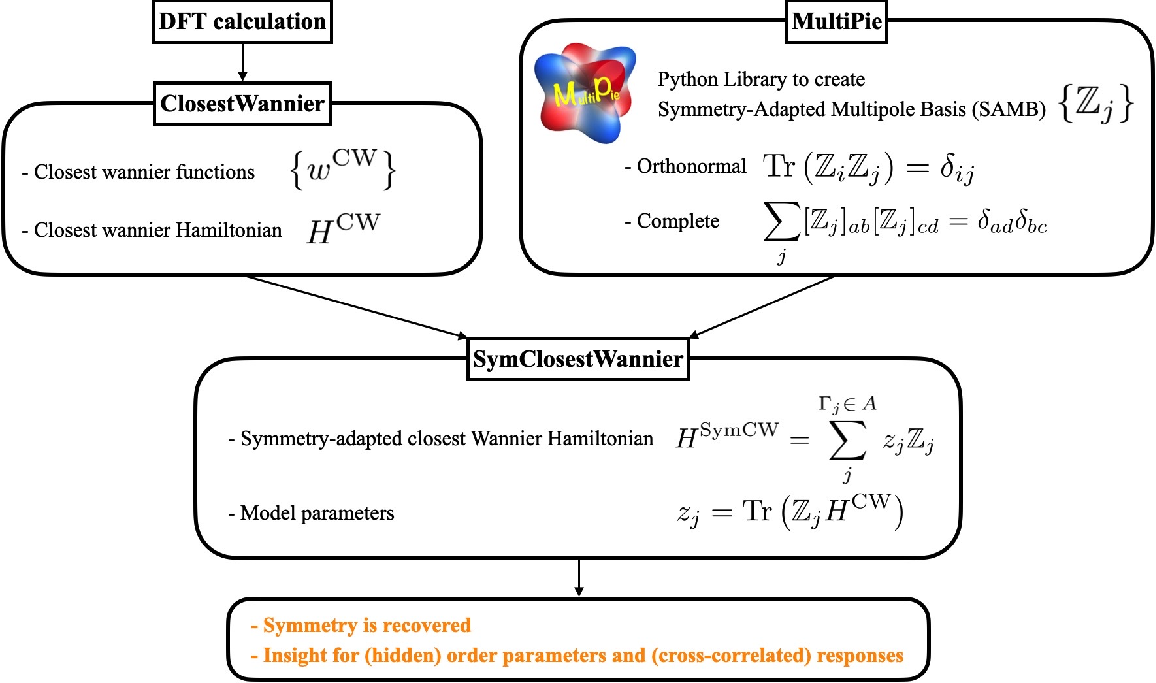}
  \caption{
  \label{fig_workflow}
  Workflow of the SymClosestWannier for constructing the symmetry-adapted closest Wannier (SymCW) model based on the symmetry-adapted multipole basis (SAMB) set generated by MultiPie (https://github.com/CMT-MU/MultiPie) and DFT input.
  }
  \end{center}
\end{figure}

In this section, we give the outline of the generation scheme of the symmetry-adapted CW model.
The workflow of our method is summarized in Fig.~\ref{fig_workflow}.

The present method consists of the following five steps:
\begin{enumerate}
 \item[(i)]
 Perform DFT calculations using DFT codes such as Quantum ESPRESSO~\cite{QE_Giannozzi_2009}, VASP~\cite{VASP_Kresse_1996}, Wien2k~\cite{Wien2k_2018}, and so on, and obtain the Kohn-Sham (KS) orbitals $\left\{\psi_{n\bm{k}}^{\rm KS}\right\}$ and KS energies $\left\{\epsilon_{n\bm{k}}^{\rm KS}\right\}$.
 \item[(ii)]
 Calculate the projection of initial atomic orbitals $\left\{\varphi_{a\alpha\bm{R}}^{\rm AO}\right\}$ onto Kohn-Sham (KS) orbitals, $\braket{\psi_{n\bm{k}}^{\rm KS}|\varphi_{a\alpha\bm{0}}^{\rm AO}}$, by using the PW2WANNIER90~\cite{QE_Giannozzi_2009}, VASP2WANNIER~\cite{VASP_Kresse_1996}, and Wien2wannier~\cite{Wien2wannier_2010} which interface Quantum ESPRESSO, VASP, and Wien2k with the Wannier90~\cite{MOSTOFI2008685, MOSTOFI20142309}.
 \item[(iii)]
 Multiply a window function $w(\epsilon)$ defined by Eq.~(\ref{eq_window}) to the projection $\braket{\psi_{n\bm{k}}^{\rm KS}|\varphi_{a\alpha\bm{0}}^{\rm AO}}$ and obtain $A(\bm{k})$ given by Eq.~(\ref{eq_proj_Amn}).
 \item[(iv)]
 Perform the singular value decomposition of $A(\bm{k})$ and obtain the unitary matrix $U(\bm{k})$.
 Then, we can construct the CWFs $\left\{w^{\rm CW}_{a\alpha\bm{R}}\right\}$ by Eq.~(\ref{eq_CWF_R}) and CW Hamiltonian $H^{\rm CW}$ by Eq.~(\ref{eq_hR}).
 \item[(v)]
 Symmetrize $H^{\rm CW}$ by expressing it as a linear combination of the SAMBs $\left\{\mathbb{Z}_{j}\right\}$ belonging to the identity irreducible representation (irrep.), $\Gamma_{j} \in A$, generated by MultiPie, $\displaystyle H^{\rm SymCW} = \sum_{j}^{\Gamma_{j} \in A} z_{j} \mathbb{Z}_{j}$. 
 Here, the coefficient for each SAMB $z_{j}$ corresponds to the model parameter, such as CEF, SOC, or real and imaginary electron hoppings.
\end{enumerate}

The steps (iii)-(v) are achieved by using ``MultiPie'' and ``SymClosestWannier'' Python packages.
Through these steps, we can symmetrize the CW Hamiltonian and obtain the model parameters $\{z_{j}\}$ with non-iterative calculations.
Note that in addition to periodic crystals, the above method can be also applied to isolated molecules by restricting the Brillouin zone sampling to the $\Gamma$ point.
As summarized in Table.~\ref{tab_samb_hamiltonian}, each SAMB corresponds to the CEF, SOC, or real and imaginary electron hoppings.
Since each SAMB is expressed as the linear combination of the direct product of the atomic and site/bond-cluster multipole basis, one can extract the electronic multipole degrees of freedom hidden in the CW Hamiltonian.
Furthermore, as summarized in Table~\ref{tab_samb_external_field_response}, our method gives us insight into various responses, because one can identify corresponding SAMB that couples with various external fields.

In order to make the present paper self-contained, the paper is organized as follows.
In Sec.~\ref{sec_CW}, we briefly review the CW formalism~\cite{Ozaki_CW_2024}.
We then explain how to construct the symmetry-adapted CW model based on SAMBs, using the monolayer graphene as an example in Sec.~\ref{sec_symcw}.
First, we show the explicit definition of SAMBs classified by D$_{\rm 6h}$ point group.
In Sec.~\ref{sec_graphene_E_field}, we demonstrate an application of our method by modeling monolayer graphene and explain how to symmetrize the CW Hamiltonian and determine the model parameters.
Additional examples are given in Supplemental Materials.
In this demonstration, we first show that under a perpendicular electric field, a slight lifting of band degeneracies on the order of $\mu$eV arises due to unexpected symmetry breaking in both the DFT calculations and the CW model. 
Subsequently, we demonstrate that our method successfully eliminates this unexpected symmetry breaking and restores the band degeneracies, allowing both qualitative and quantitative analysis of the band structure under the correct symmetry.
We also discuss possible unusual responses predicted from the obtained results.
The last section summarizes the paper.

\section{Closest Wannier formalism}
\label{sec_CW}

\begin{figure*}[t]
  \begin{center}
  \includegraphics[width=0.98 \hsize]{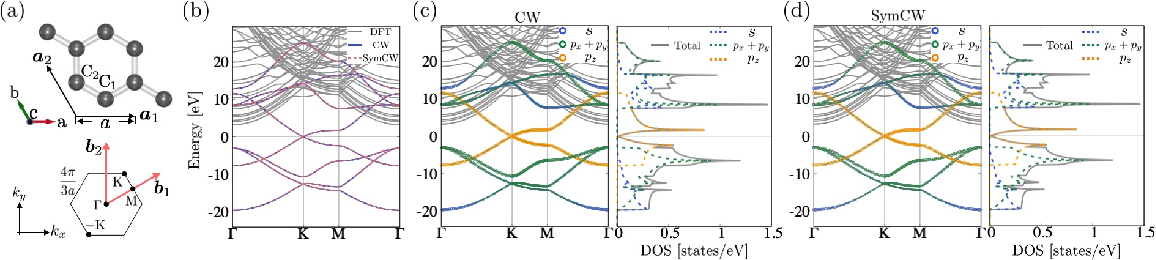}
  \vspace{5mm}
  \caption{
  \label{fig_graphene_cw_band}
  The comparisons of energy dispersions among the DFT calculation, CW model, and the symmetry-adapted CW (SymCW) model.
  (a) Crystal structure of monolayer graphene and the Brillouin zone and high-symmetry points.
  (b) The comparison between the band dispersions obtained by the DFT calculation (gray solid lines), CW model (blue solid lines), and the SymCW model (red dashed lines).
  The orbital dependence of the band dispersion and DOS obtained by (c) CW and (d) SymCW models, where the blue, green, and orange lines represent the CWFs with $s$, $p_{x}$, $p_{y}$, and $p_{z}$ orbital-like symmetry, as shown in Fig.~\ref{fig_graphene_cw}.
  The Fermi energy is set as the origin.
  All figures for crystal structures and wave functions in this paper are drawn by QtDraw~\cite{PhysRevB.107.195118}.
  }
  \end{center}
\end{figure*}

In this section, we briefly review the closest Wannier (CW) formalism developed by T. Ozaki~\cite{Ozaki_CW_2024}.
Following the proposed method~\cite{Ozaki_CW_2024}, we can obtain a Wannier function that is the ``closest'' to an original atomic orbital in a Hilbert space with non-iterative calculations, as shown in the following subsections, Sec.~\ref{sec_proj} and Sec.~\ref{sec_CW_model}.

\subsection{Projection of atomic orbitals onto Kohn-Sham orbitals}
\label{sec_proj}

Let us begin with a KS equation~\cite{Kohn_Sham_1965} based on DFT~\cite{Hohenberg_Kohn_1964}:
\begin{align}
H^{\rm KS} \ket{\psi_{n\bm{k}}^{\rm KS}} = \epsilon_{n\bm{k}}^{\rm KS} \ket{\psi_{n\bm{k}}^{\rm KS}},
\end{align}
where $H^{\rm KS}$ is the KS Hamiltonian, and $\psi_{n\bm{k}}^{\rm KS}$ ($\epsilon_{n\bm{k}}^{\rm KS}$) represents the $n$th KS orbital (energy) at a crystal momentum $\bm{k}$.
$N_{\rm B}$ KS orbitals are orthonormalized as $\braket{\psi_{n\bm{k}}^{\rm KS}|\psi_{n'\bm{k}'}^{\rm KS}} = \delta_{\bm{k},\bm{k}'} \delta_{n,n'}$.

As an initial guess for WFs, we consider $N_{\rm W}$ $(\leq N_{\rm B})$ hydrogenic atomic orbitals (AOs), such as those implemented in Wannier90 given by
\begin{align}
\varphi_{a\alpha\bm{R}}^{\rm AO}(\bm{r}) \equiv \varphi_{nlm\sigma}^{\rm AO}(\bm{r} - \bm{R} - \bm{\alpha}),
\label{eq_AO}
\end{align}
where an abbreviation $a = (n, l, m, \sigma)$ is introduced, and $l$, $m$, $\sigma$, and $n$ denote the azimuthal quantum number, magnetic quantum number or a subscript of the orbitals in real representation, spin, and the other quantum number, such as the principal quantum number, respectively.
$\bm{R}$ and $\bm{\alpha}$ denote the lattice vector and the position of the sublattice $\alpha$ within the home-unit-cell, $\bm{R} = \bm{0}$.

A main idea of the CW formalism is to introduce a window function $w (\epsilon)$ for the projection of the AO onto KS orbitals as
\begin{align}
&
\ket{\phi_{a\alpha\bm{R}}^{\rm AO}}=\frac{1}{\sqrt{N_{k}}} \sum_{n\bm{k}} e^{-i \bm{k} \cdot (\bm{R} + \bm{\alpha})} \ket{\psi_{n\bm{k}}^{\rm KS}} A_{n, a\alpha}^{}(\bm{k}),
\label{eq_proj_AO}
\\ &
A_{n, a\alpha}^{}(\bm{k}) = w (\epsilon_{n\bm{k}}^{\rm KS}) \braket{\psi_{n\bm{k}}^{\rm KS}|\varphi_{a\alpha\bm{0}}^{\rm AO}},
\label{eq_proj_Amn}
\end{align}
where $N_{k}$ is the number of $\bm{k}$ points and the $N_{\rm B} \times N_{\rm W}$ projection matrix $A(\bm{k})$ is weighted with a window function $w (\epsilon)$:
\begin{align}
&
w (\epsilon) = f(x_{0}) + f(x_{1}) - 1 + \delta,\quad x_{0} = \frac{\epsilon_{0} - \epsilon}{k_{\rm B} T_{0}},\quad x_{1} = \frac{\epsilon - \epsilon_{1}}{k_{\rm B} T_{1}}.
\label{eq_window}
\end{align}
Here
\begin{align}
f(x) = 
\begin{cases}
1/(\exp(x) + 1) & (T > 0) \\
1 \, (x < 0), 0 \, (x \geq 0) & (T = 0) \\
\end{cases}
\end{align}
denotes the Fermi-Dirac distribution function.
While the standard disentanglement approach is required to specify the outer- and frozen-windows, the CW formalism only needs one window defined by $\epsilon_{0}$ and $\epsilon_{1}$ ($\epsilon_{0} < \epsilon_{1}$) representing the lower and upper edge of the energy window, while $T_{0}$ and $T_{1}$ are introduced to control the degree of smearing around $\epsilon_{0}$ and $\epsilon_{1}$, respectively.
As a result of this smearing, all KS orbitals are included in the projection given by Eq.~(\ref{eq_proj_AO}) with weights, where the weight is particularly large inside the window ($\epsilon_{0} < \epsilon_{n\bm{k}}^{\rm KS} < \epsilon_{1}$), whereas it is small outside of the window.
In Eq.~(\ref{eq_window}), $\delta$ is a small constant (e.g., $10^{-12}$) introduced to prevent the matrix consisting of $A_{n, a\alpha}^{}(\bm{k})$ from becoming ill-conditioned~\cite{Ozaki_CW_2024}.

The Bloch representation of the atomic orbital is given by the Fourier transform of Eq.~(\ref{eq_proj_AO}):
\begin{align}
\ket{\phi_{a\alpha\bm{k}}^{\rm AO}}=\frac{1}{\sqrt{N_{k}}}\sum_{\bm{R}} e^{i \bm{k} \cdot (\bm{R}+\bm{\alpha})} \ket{\phi_{a\alpha\bm{R}}^{\rm AO}}=\sum_{n} \ket{\psi_{n\bm{k}}^{\rm KS}} A_{n, a\alpha}^{}(\bm{k}).
\end{align}
Since the AOs are not orthonormal, the $N_{\rm W} \times N_{\rm W}$ weighted overlap matrix $S^{}(\bm{k})$ is not an identity matrix and given by
\begin{align}
& S_{a\alpha, a'\alpha'}(\bm{k}) = \braket{\phi_{a\alpha\bm{k}}^{\rm AO}|\phi_{a'\alpha'\bm{k}}^{\rm AO}} = \sum_{n} A_{n,a\alpha}^{*}(\bm{k}) A_{n, a'\alpha'}^{}(\bm{k}),
\label{eq_S_k}
\\ &
S_{a\alpha, a'\alpha'}(\bm{R}) = \braket{\phi_{a\alpha\bm{0}}^{\rm AO}|\phi_{a'\alpha'\bm{R}}^{\rm AO}}
= \frac{1}{N_{k}} \sum_{\bm{k}} e^{-i\bm{k}\cdot(\bm{R}+\bm{\alpha}' - \bm{\alpha})} S_{a\alpha, a'\alpha'}(\bm{k}).
\label{eq_S_R}
\end{align}

\subsection{Construction of Closest Wannier model}
\label{sec_CW_model}

The orthonormal WFs to be optimized are defined by a linear combination of the KS orbitals as
\begin{align}
&
\ket{w_{a\alpha\bm{R}}^{}}=\frac{1}{\sqrt{N_{k}}} \sum_{n\bm{k}} e^{-i \bm{k} \cdot (\bm{R} + \bm{\alpha})}
\ket{\psi_{n\bm{k}}^{\rm KS}} B_{n, a\alpha}^{}(\bm{k}),
\end{align}
where $B(\bm{k})$ is the $N_{\rm B} \times N_{\rm W}$ partial unitary matrix satisfying $B^{\dagger}(\bm{k})B^{}(\bm{k}) = I_{N_{\rm W} \times N_{\rm W}}$ ($I_{N_{\rm W} \times N_{\rm W}}$ is an $N_{\rm W} \times N_{\rm W}$ identity matrix).
Different from the conventional Maximally localized Wannier functions (MLWFs)~\cite{PhysRevB.56.12847, PhysRevB.65.035109,RevModPhys.84.1419,MOSTOFI2008685,MOSTOFI20142309,Pizzi_2020}, where the MLWFs are obtained by minimizing the spread of the WFs, the CWFs are obtained by minimizing the distance measure function $F[B]$ defined by the sum of inner product of the difference between the WFs and their corresponding AOs used as an initial guess~\cite{Ozaki_CW_2024}:
\begin{align}
F[B] &= \sum_{a\alpha} \left(\bra{\phi_{a\alpha\bm{R}}^{\rm AO}} - \bra{w_{a\alpha\bm{R}}^{}}\right)\left(\ket{\phi_{a\alpha\bm{R}}^{\rm AO}} - \ket{w_{a\alpha\bm{R}}^{}}\right),
\cr &
= \frac{1}{N_{k}} \sum_{\bm{k}} \mathrm{Tr}\left[ (A^{\dagger}(\bm{k}) - B^{\dagger}(\bm{k}))(A^{}(\bm{k}) - B^{}(\bm{k})) \right].
\end{align}
Then, the minimum of $F[B]$ is realized when $B^{}(\bm{k}) = U^{}(\bm{k})$, where $U^{}(\bm{k})$ is defined by
\begin{align}
U^{}(\bm{k}) = W^{}(\bm{k}) V^{\dagger}(\bm{k}),
\label{eq_U_SVD}
\end{align}
Here, $W(\bm{k})$ and $V(\bm{k})$ are obtained by the singular value decomposition (SVD) of $A^{}(\bm{k})$:
\begin{align}
A^{} = W^{} \Sigma^{} V^{\dagger} = W^{} V^{\dagger} V^{} \Sigma^{} V^{\dagger} = U P,
\label{eq_A_SVD}
\end{align}
where we have omitted the $\bm{k}$ dependence for notational simplicity.
$\Sigma$ is the $N_{\rm W} \times N_{\rm W}$ singular value diagonal matrix, and $ W^{}$ ($V^{}$) is the $N_{\rm B} \times N_{\rm W}$ ($N_{\rm W} \times N_{\rm W}$) left (right) singular matrix.
$W^{}$ is the partial unitary matrix; $W^{\dagger} W^{} = I $ ($W^{} W^{\dagger} \neq I $) and $V^{}$ is a full unitary matrix; $V^{\dagger} V^{} = V^{} V^{\dagger} = I$.
Using $\Sigma^{}$, $W^{}$, and $V^{}$, the weighted overlap matrix $S^{}(\bm{k})$ is expressed by
\begin{align}
S(\bm{k}) = V(\bm{k}) \Sigma^{2}(\bm{k}) V^{\dagger}(\bm{k}) = P^{2}(\bm{k}).
\label{eq_S_SVD}
\end{align}
When $P(\bm{k})$ is invertible, combining the Eqs.~(\ref{eq_A_SVD}) and (\ref{eq_S_SVD}), $U^{}(\bm{k})$ can be expressed as
\begin{align}
U^{}(\bm{k}) = A^{}(\bm{k}) S^{-1/2}(\bm{k}).
\label{eq_U_Lowdin}
\end{align}
Equation.~(\ref{eq_U_Lowdin}) is nothing but the partial unitary matrix defined by the L\"owdin orthonormalization~\cite{Lowdin_1950}, which is consistent with the fact that the closest property of the L\"owdin orthogonalized orbitals to a given set of AOs~\cite{Carlson_1957}.
Using $U^{}(\bm{k})$, the CWF is obtained by
\begin{align}
&
\ket{w_{a\alpha\bm{R}}^{\rm CW}}=\frac{1}{\sqrt{N_{k}}} \sum_{n\bm{k}} e^{-i \bm{k} \cdot (\bm{R} + \bm{\alpha})}
\ket{\psi_{n\bm{k}}^{\rm KS}} U_{n, a\alpha}(\bm{k}),
\label{eq_CWF_R}
\\ &
\ket{w_{a\alpha\bm{k}}^{\rm CW}} = \frac{1}{\sqrt{N_{k}}} \sum_{\bm{R}} e^{i \bm{k} \cdot (\bm{R}+\bm{\alpha})} \ket{w_{a\alpha\bm{R}}^{\rm CW}}  = \sum_{a'\alpha'} \ket{\phi_{a'\alpha'\bm{k}}^{\rm AO}} S_{a'\alpha',a\alpha}^{-1/2}(\bm{k}).
\label{eq_CWF_k}
\end{align}
Since Eq.~(\ref{eq_CWF_R}) minimizes $F[B]$, this is called as the ``closest'' Wannier function to an initial AO in a Hilbert space.
Notably, no iterative calculations are required for minimizing $F[B]$ in this formalism,
Therefore, the CWFs can be constructed with no iterative calculations as long as the projection matrix $A(\bm{k})$ is positive definite, significantly reducing the computational costs.
Furthermore, the disentanglement of bands is naturally taken into account by introducing a window function given by Eq.~(\ref{eq_window})~\cite{Ozaki_CW_2024}.

\begin{figure*}[t]
  \begin{center}
  \includegraphics[width=0.98 \hsize]{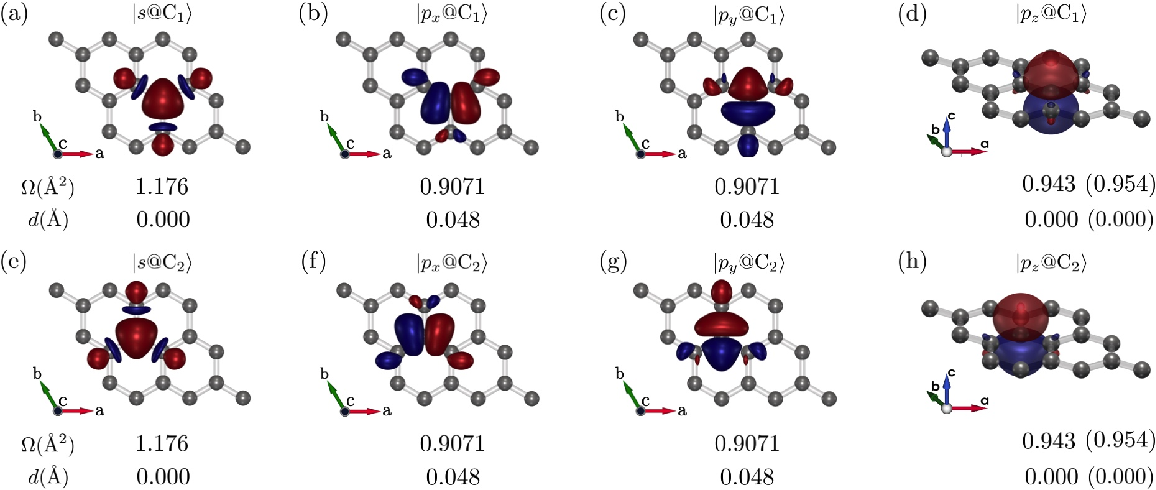}
  \vspace{5mm}
  \caption{
  \label{fig_graphene_cw}
  The CWFs with $s$, $p_{x}$, $p_{y}$, and $p_{z}$ like symmetry.
  The upper and lower panels represent the CWFs localized at around C$_{1}$ and C$_{2}$ sites, respectively.
  $\Omega (\AA^{2})$ and $d(\AA)$ denote the spread of the CWFs and the distance between the center of the CWFs and the nearest-neighbor atom, where the values in parentheses in (d) and (h) are from Ref.~\cite{qiao_proj_2023}.
  }
  \end{center}
\end{figure*}

Once $U^{}(\bm{k})$ is obtained by Eq.~(\ref{eq_U_SVD}) or (\ref{eq_U_Lowdin}), the CW Hamiltonian is obtained from the KS energies as
\begin{align}
H^{\rm CW}_{a,a'}(\bm{b}@\bm{c})
&= \frac{1}{N_{k}} \sum_{\bm{k}} \sum_{n} e^{i\bm{k}\cdot\bm{b}} U_{n, a\alpha}^{*}(\bm{k}) \epsilon_{n\bm{k}}^{\rm KS} U_{n, a'\alpha'}^{}(\bm{k}),
\label{eq_hR}
\\
H^{\rm CW}_{a\alpha,a'\alpha'}(\bm{k}) &= \sum_{\bm{R}} e^{-i\bm{k}\cdot\bm{b}} H^{\rm CW}_{a\alpha,a'\alpha'}(\bm{b}@\bm{c}),
\label{eq_hk}
\end{align}
where we have introduced a bond-vector $\bm{b} = (\bm{R}+\bm{\alpha}) - (\bm{R}'+\bm{\alpha}')$ and bond-center $\bm{c} = [(\bm{R}+\bm{\alpha}) + (\bm{R}'+\bm{\alpha}')]/2$.

Taking monolayer graphene as an example, the band dispersion obtained from the CW model is shown in Fig.~\ref{fig_graphene_cw_band}.
As shown in Fig.~\ref{fig_graphene_cw_band}(a), the lattice constant is $a = 2.456\, \AA$, the length of the vacuum layer along the $c$ axis is set as $c = 20\, \AA$, and the unit vectors are given by $\bm{a}_{1} = (1,0,0) a$, $\bm{a}_{2} = (-1/2,\sqrt{3}/2,0) a$, and $\bm{a}_{3} = (0,0,1) c$.
As shown in Fig.~\ref{fig_graphene_cw_band}(b), since the bands near the Fermi energy are strongly entangled with many free-electron bands, we set the energy window as $[\epsilon_{0}, \epsilon_{1}] = [-20,-5]$ and use a relatively large smearing, $[T_{0}, T_{1}] = [0.0,6.0]$ in order to properly disentangle the bands.
As for the initial guesses, we have chosen hydrogenic $s$, $p_{x}$, $p_{y}$, and $p_{z}$ orbitals for each C atom.
Then, by using SymClosestWannier, we obtain the $s$, $p_{x}$, $p_{y}$, and $p_{z}$ like CWFs as depicted in Fig.~\ref{fig_graphene_cw}.
The obtained band dispersions are indicated by blue solid lines as shown in Fig.~\ref{fig_graphene_cw_band}(b), reproducing the DFT band dispersion indicated by gray solid lines with high accuracy.

The upper and lower panels of Fig.~\ref{fig_graphene_cw} represent the $s$, $p_{x}$, $p_{y}$, and $p_{z}$ like CWFs localized at around C$_{1}$ and C$_{2}$ sites, where $\Omega (\AA^{2})$ and $d(\AA)$ denote the spread of the CWFs and the distance between the center of the CWFs from the nearest-neighbor atom.
The values in parentheses in Figs.~\ref{fig_graphene_cw}(d) and (h) are those of the MLWFs obtained from the hydrogenic $p_{z}$ orbital based on the standard approach~\cite{qiao_proj_2023}.
The localization of $p_{z}$ like CWF is almost equivalent to that of MLWF, while the spread of the $s$, $p_{x}$, and $p_{y}$ like CWFs are larger than that of the other three MLWFs corresponding to the hybridized $s \pm (p_{x},p_{y})$ orbitals with $\Omega = 0.446~\AA^{2}$~\cite{qiao_proj_2023}.
On the other hand, due to the closest property of the CWFs to the original AOs, the obtained CWFs are well-localized at each C atom, and $d(\AA)$ of the $p_{x}$ and $p_{y}$ like CWFs are much smaller than that of the hybridized $s \pm (p_{x},p_{y})$ like MLWFs with $d = 0.394~\AA$~\cite{qiao_proj_2023}.
Note that although the $s$, $p_{x}$, and $p_{y}$ like CWFs are distorted from the original shape, their symmetry properties are equivalent to that of the original ones as discussed in the next section.

\subsection{Symmetry properties of the closest Wannier functions and Hamiltonian}
\label{sec_cw_symmetry}

Let us consider the symmetry properties of the CWFs and Hamiltonian.
First of all, within the space group $G$ of the given system, the CW Hamiltonian must be invariant under any symmetry operations, $g = \left\{p|\bm{t}\right\} \in G$ ($p$ and $\bm{t}$ are the rotation/inversion and fractional translation, respectively):
\begin{align}
  gH^{\rm CW} g^{-1} = H^{\rm CW}.
 \label{eq_H_sym}
\end{align}
However, Eq.~(\ref{eq_H_sym}) is not often fully satisfied because no symmetry constraint is applied during the wannierization process, leading to unexpected symmetry breaking and slight lifting of band degeneracies~\cite{PhysRevMaterials.2.103805}.

The AO is transformed by $g$ as
\begin{align}
g \ket{\phi_{a\alpha\bm{R}}^{\rm AO}} = \sum_{a'\alpha'\bm{R}'} \ket{\phi_{a'\alpha'\bm{R}'}^{\rm AO}} U_{a'\alpha'\bm{R}', a\alpha\bm{R}}^{}(g),
\end{align}
where $U(g)$ is a $N_{\rm W} \times N_{\rm W}$ transformation matrix that can be decomposed into the product of the site part $R^{\rm (s)}(g)$ and the atomic part $R^{\rm (a)}(p)$ as
\begin{align}
&
U_{a'\alpha'\bm{R}', a\alpha\bm{R}}^{}(g) = R_{\alpha'\bm{R}',\alpha\bm{R}}^{\rm (s)}(g) R_{a',a}^{\rm (a)}(p),
\label{eq_sym_oper_AO}
\\ &
R_{\alpha'\bm{R}',\alpha\bm{R}}^{\rm (s)}(g) = \delta_{\bm{R}'+\bm{\alpha}', g(\bm{R}+\bm{\alpha})}.
\label{eq_sym_oper_AO_R}
\end{align}
The Bloch representation of the AO is transformed by $g$ as
\begin{align}
&
g \ket{\phi_{a\alpha\bm{k}}^{\rm AO}} = \sum_{a'\alpha'} \ket{\phi_{a'\alpha' p\bm{k}}^{\rm AO}}  U_{a'\alpha', a\alpha}(g; \bm{k}),
\label{eq_gAO}
\\ &
U(g; \bm{k}) =  e^{-ip\bm{k} \cdot \bm{t}} U(g).
\label{eq_sym_oper_AO_k}
\end{align}
Note that $U(g)$ is the $\bm{k}$ independent transformation matrix~\cite{PhysRevMaterials.2.103805}:
\begin{align}
&
U_{a'\alpha', a\alpha}(g) = R_{\alpha',\alpha}^{\rm (s)'}(g) R_{a',a}^{\rm (a)}(p),
\label{eq_Ug}
\\ &
R_{\alpha',\alpha}^{\rm (s)'}(g) =
\begin{cases} 1 & g \bm{\alpha} - \bm{\alpha}' \in \text{lattice vector}, \\ 0 & \text{otherwise}, \end{cases}
\end{align}

The overlap matrix satisfies the following relation
\begin{align}
S(p\bm{k}) = U^{}(g;\bm{k}) S(\bm{k}) U^{\dagger}(g;\bm{k}) = U^{}(g) S(\bm{k}) U^{\dagger}(g),
\end{align}
Considering $S^{n}(p\bm{k}) = U^{}(g) S^{n}(\bm{k}) U^{\dagger}(g)$, $S^{-1/2}(\bm{k})$ satisfies
\begin{align}
S^{-1/2}(p\bm{k}) = U^{}(g) S^{-1/2}(\bm{k}) U^{\dagger}(g).
\label{eq_gSk}
\end{align}
Using Eqs.~(\ref{eq_gAO}) and (\ref{eq_gSk}), the CWFs are transformed by $g$ as
\begin{align}
g \ket{w_{a\alpha\bm{k}}^{\rm CW}} = \sum_{a'\alpha'} \ket{w_{a'\alpha' p\bm{k}}^{\rm CW}} U_{a'\alpha', a\alpha}(g; \bm{k}).
\label{eq_gCW_k}
\end{align}
Thus, the CWFs are transformed in the same way as the AOs given by Eq.~(\ref{eq_gAO}) because the CWFs are obtained by the symmetric L\"owdin orthonormalization of a given set of AOs~\cite{Lowdin_1950}.
Similarly, the real space representation of the CWFs are transformed in the same way as the AOs:
\begin{align}
g \ket{w_{a\alpha\bm{R}}^{\rm CW}} = \sum_{a'\alpha'\bm{R}'} \ket{w_{a'\alpha'\bm{R}'}^{\rm CW}} U_{a'\alpha'\bm{R}', a\alpha\bm{R}}^{}(g).
\label{eq_gCW_R}
\end{align}

Although the shape of the CWFs is distorted from the original AOs and the ($l$, $m$) of the original AOs are no longer strictly preserved in the CWFs as shown in Fig.~\ref{fig_graphene_cw}, however, as indicated in Eq.~(\ref{eq_gCW_R}), the symmetry operations are inherited from the original AOs.
Then, as demonstrated by D. Gresch {\it et al.}~\cite{PhysRevMaterials.2.103805}, $H^{\rm CW}$ can be symmetrized by using the same transformation matrix $U(g)$ as for the AOs:
\begin{align}
  H^{\rm CW} \rightarrow \frac{1}{N_{g}} \sum_{g \in G} U(g) H^{\rm CW} U(g^{-1}),
\end{align}
where $N_{g}$ is the number of symmetry operations.

On the other hand, in this paper, we propose another method that not only symmetrizes the CW Hamiltonian but also provides a physical interpretation, following the approach outlined below.
First, suppose that we have a complete orthonormal basis set of matrices $\left\{\mathbb{Z}_{j}\right\}$ defined in the Hilbert space of the CWFs, satisfying the orthonormal and complete relations as
\begin{align}
  &
  \sum_{j} \braket{w^{\rm CW}_{a_{1}\alpha_{1}\bm{R}_{1}}|\mathbb{Z}_{j}|w^{\rm CW}_{a_{2}\alpha_{2}\bm{R}_{2}}} \braket{w^{\rm CW}_{a_{3}\alpha_{3}\bm{R}_{3}}|\mathbb{Z}_{j}|w^{\rm CW}_{a_{4}\alpha_{4}\bm{R}_{4}}} = \delta_{1,4} \delta_{2,3},
  \label{eq_c_samb_R_comp_}
  \\ &
  \mathrm{Tr}\left[\mathbb{Z}_{i} \mathbb{Z}_{j}\right] = \delta_{i,j}.
  \label{eq_c_samb_R_ortho_}
\end{align}
Second, suppose that $\mathbb{Z}_{j}$ is classified by the irrep. $\Gamma$ and its component $\gamma$ of the point group associated with the space group.
Since $H^{\rm CW}$ must be invariant for all the symmetry operations as given by Eq.~(\ref{eq_H_sym}), we can express $H^{\rm CW}$ as a linear combination of $\mathbb{Z}_{j}$ belonging to the identity irrep. ($A$) satisfying $g \mathbb{Z}_{j} g^{-1} = \mathbb{Z}_{j}$ as
\begin{align}
  H^{\rm SymCW} = \sum_{j}^{\Gamma_{j} \in A} z_{j} \mathbb{Z}_{j}.
  \label{eq_Hcw_samb_}
\end{align}
where $\{z_{j}\}$ are the coefficients for each SAMB corresponding to the model parameters.
Here $z_{j}$ is determined by the following simple matrix projection as
\begin{align}
  z_{j} = \mathrm{Tr}\left[ \mathbb{Z}_{j} H^{\rm CW} \right].
\end{align}
In addition, when the system preserves the time-reversal symmetry $\mathbb{Z}_{j}$ must be time-reversal even, otherwise, $\mathbb{Z}_{j}$ can be time-reversal odd.
Since each $\mathbb{Z}_{j}$ in Eq.~(\ref{eq_Hcw_samb_}) is fully symmetric, the Hamiltonian $H^{\rm SymCW}$ becomes fully symmetric as well.

In order to realize the above method, $\mathbb{Z}_{j}$ is needed to be characterized by the spatial-inversion and time-reversal parities, and the irrep. of the point group.
In this way, the symmetry-adapted multipole basis~\cite{JPSJ.87.033709, PhysRevB.98.165110, PhysRevB.102.144441, JPSJ.89.104704} is a suitable description for $\mathbb{Z}_{j}$.
In the next section, we will outline how to systematically generate such a basis set and to construct a symmetry-adapted CW model.

\section{Symmetry-adapted closest Wannier modeling}
\label{sec_symcw}

\begin{table}
  \renewcommand{\arraystretch}{1.2}
  \begin{center}
  \caption{ \label{tab_four_samb}
  Four types of SAMB classified according to the spatial-inversion ($\mathcal{P}$) and time-reversal ($\mathcal{T}$) parities.
  }
  \begin{ruledtabular}
  \begin{tabular}{llll}
  Type                   & Symbol & $\mathcal{P}$ & $\mathcal{T}$
  \\ \hline
  Electric (E)           & $\mathbb{Q}_{lm}, \mathbb{Q}_{l\xi}$ & $(-1)^{l}$ (polar) & $+$ \\
  Magnetic (M)           & $\mathbb{M}_{lm}, \mathbb{M}_{l\xi}$ & $(-1)^{l+1}$ (axial) & $-$ \\
  Magnetic toroidal (MT) & $\mathbb{T}_{lm}, \mathbb{T}_{l\xi}$ & $(-1)^{l}$ (polar) & $-$ \\
  Electric toroidal (ET) & $\mathbb{G}_{lm}, \mathbb{G}_{l\xi}$ & $(-1)^{l+1}$ (axial) & $+$ \\
\end{tabular}
\end{ruledtabular}
\end{center}
\end{table}

Previous studies have developed a systematic method for generating a set of SAMBs~\cite{JPSJ.87.033709, PhysRevB.98.165110, PhysRevB.102.144441, JPSJ.89.104704}.
The SAMBs are classified into four types of multipoles, E, M, MT, and ET multipoles, according to the spatial-inversion ($\mathcal{P}$) and time-reversal ($\mathcal{T}$) parities, as summarized in Table~\ref{tab_four_samb}.
Although SAMBs were originally defined in the Hilbert space of pure AOs, they can also be defined in the Hilbert space of CWFs, as the symmetry properties of CWFs are common to those of AOs.
In this case, the SAMBs retain the same $\mathcal{P}$ and $\mathcal{T}$ parities, as well as the irreducible representations and their components, as those defined for pure AOs.
It should be noted that while the rank of the multipoles is no longer strictly preserved, it is partially inherited from the original definition due to the close similarity between the CWFs and AOs.

In the following subsections, we provide the explicit definition of SAMBs and demonstrate the generation scheme of the symmetry-adapted CW (SymCW) model.
As indicated in Eq.~(\ref{eq_sym_oper_AO}), the symmetry operations can be applied separately to atomic positions and the internal atomic degrees of freedom.
Given this separable nature, SAMBs can be also constructed independently for atomic sites and bonds, as well as for the internal atomic degrees of freedom~\cite{PhysRevB.107.195118}.
Then, we introduce the SAMB for the atomic degrees of freedom, referred to as ``atomic SAMB,'' and the SAMB for site/bond degrees of freedom, called as ``cluster SAMB.''
By performing the irreducible decomposition of the direct product of the atomic SAMB and the cluster SAMB, we obtain the SAMB for the total Hilbert space, which we term ``combined SAMB.''
Taking monolayer graphene as a case study, we show the explicit definition of the cluster SAMB in Sec.~\ref{sec_cluster_samb} and the atomic SAMB in Sec.~\ref{sec_atomic_samb}.
The combined SAMB is introduced in Sec.~\ref{sec_combined_samb}.
Then, the generation scheme of the SymCW model is shown in Sec.~\ref{sec_symcw_ham}.

\subsection{Symmetry-adapted harmonics}
\label{sec_harmonics}

\begin{table}[t]
\renewcommand{\arraystretch}{1.2}
\begin{center}
\caption{ \label{tab_harmonics}
Polar and axial harmonics up to rank 3 in D$_{\rm 6h}$ point group.
$r = \sqrt{x^{2}+y^{2}+z^{2}}$ and $R=\sqrt{X^{2}+Y^{2}+Z^{2}}$.
The label $g$ and $u$ are exchanged in the irrep. of the axial form.
The multiplicity $n$ is omitted for simplicity.
}
\begin{ruledtabular}
\begin{tabular}{ccccc}
$l$ & $\Gamma$    & $\gamma$ & Polar form & Axial form \\
\hline
0    & $A_{1g}$       & -                 & $1$       & $1$      \\
\hline
1    & $A_{2u}$       & -                 & $z$       & $Z$      \\
      & $E_{1u}$       & $u$                & $x$       & $-Y$     \\
      &                       & $v$                & $y$       & $X$      \\
\hline
2    & $A_{1g}$       & -                 & $\frac{1}{2}\left(3z^{2}-r^{2}\right)$       & $\frac{1}{2}\left(3Z^{2}-R^{2}\right)$    \\
      & $E_{1g}$       & $u$                & $\sqrt{3}zx$       & $-\sqrt{3}YZ$     \\
      &                       & $v$                & $\sqrt{3}yz$       & $\sqrt{3}ZX$      \\
      & $E_{2g}$       & $u$                & $\frac{\sqrt{3}}{2}\left(x^{2}-y^{2}\right)$       & $\sqrt{3}XY$     \\
      &                       & $v$                & $-\sqrt{3}xy$       & $\frac{\sqrt{3}}{2}\left(X^{2}-Y^{2}\right)$    \\
\hline
3    & $A_{2u}$       & -                 & $\frac{1}{2} z\left(5z^{2} - 3r^{2}\right) $       & $\frac{1}{2} Z\left(5Z^{2} - 3R^{2}\right) $ \\
      & $B_{1u}$       & -                 & $\frac{\sqrt{10}}{4}y\left(3x^{2}-y^{2}\right)$       &  $\frac{\sqrt{10}}{4}Y\left(3X^{2}-Y^{2}\right)$  \\
      & $B_{2u}$       & -                 & $\frac{\sqrt{10}}{4}x\left(x^{2}-3y^{2}\right)$       &  $\frac{\sqrt{10}}{4}X\left(X^{2}-3Y^{2}\right)$  \\
      & $E_{1u}$       & $u$                & $\frac{\sqrt{6}}{4}x\left(5z^{2}-r^{2}\right)$       &  $-\frac{\sqrt{6}}{4}Y\left(5Z^{2}-R^{2}\right)$     \\
      &                       & $v$                & $\frac{\sqrt{6}}{4}y\left(5z^{2}-r^{2}\right)$       & $\frac{\sqrt{6}}{4}X\left(5Z^{2}-R^{2}\right)$    \\
      & $E_{2u}$       & $u$                & $\frac{\sqrt{15}}{2}z\left(x^{2}-y^{2}\right)$       & $\sqrt{15}XYZ$     \\
      &                       & $v$                & $-\sqrt{15}xyz$       & $\frac{\sqrt{15}}{2}Z\left(X^{2}-Y^{2}\right)$    \
\end{tabular}
\end{ruledtabular}
\end{center}
\end{table}

Let us begin with the normalized symmetry-adapted harmonics $O_{l\xi}$ ($\xi = (\Gamma,n,\gamma)$) defined by
\begin{align}
&
O_{l\xi}(\bm{r})=\sum_{m}U_{m,\xi}^{(l)}O_{lm}(\bm{r}),
\label{eq_Olxi}
\\ &
O_{lm}(\bm{r}) = r^{l} \sqrt{\frac{4\pi}{2l+1}} Y_{lm}(\hat{\bm{r}}),\quad r = |\bm{r}|, \quad \hat{\bm{r}} = \frac{\bm{r}}{r},
\label{eq_Olm}
\end{align}
where $O_{lm}$ is the normalized spherical harmonics proportional to the spherical harmonics $Y_{lm}$ with the rank $l$ and component $-l \leq m \leq l$.
$U^{(l)}$ is an unitary matrix for basis transformation from $(l,m)$ basis of the rotation group to $(l,\xi)$ basis of the point group.
$\Gamma$ and $\gamma$ denote the irrep. and its component, respectively, and the label $n$ represents the multiplicity to distinguish independent harmonics belonging to the same irrep.
For example, the harmonics up to rank 3 in the D$_{\rm 6h}$ point group are given in Table~\ref{tab_harmonics}.

\subsection{Cluster SAMB}
\label{sec_cluster_samb}

\begin{figure}[t]
  \begin{center}
  \includegraphics[width=0.95 \hsize]{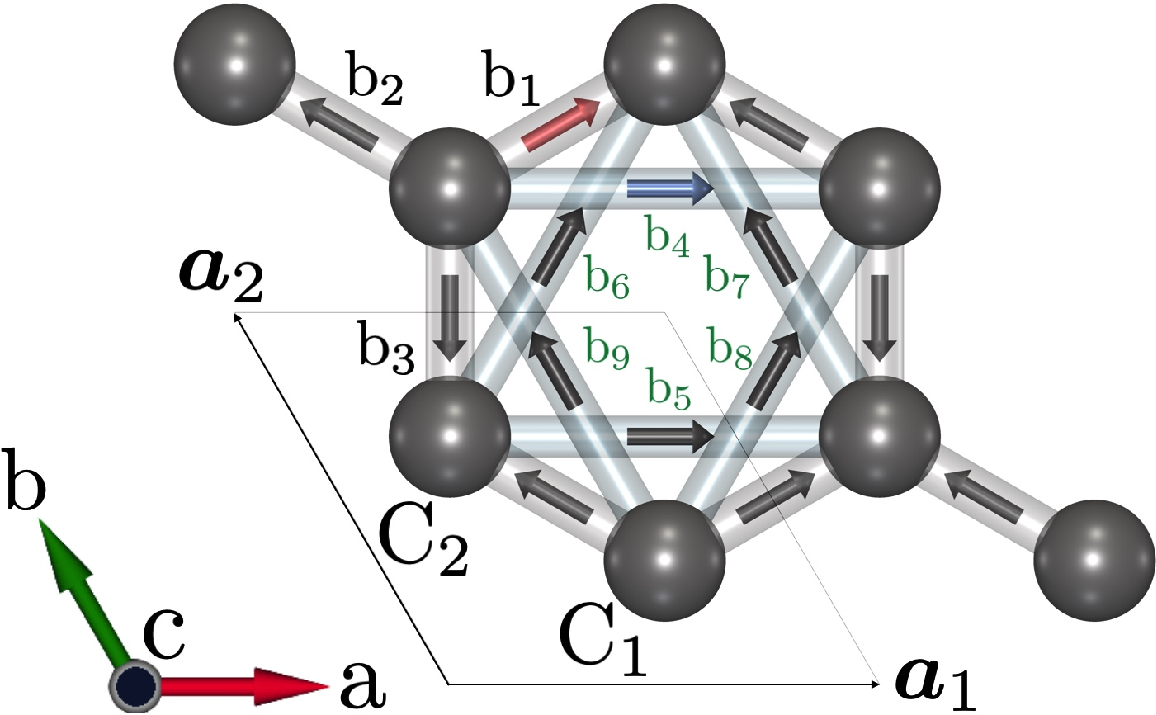}
  \vspace{5mm}
  \caption{
  \label{fig_sb_cluster}
Site-cluster, and the nearest-neighbor and 2nd-neighbor bond-clusters in graphene, where the red and blue arrows stand for the representative bonds.
  }
  \end{center}
\end{figure}

\begin{table}[t]
  \caption{\label{tbl_graphene_sb}
  Site-cluster (C) and nearest-neighbor (B$_{1}$) and 2nd-neighbor (B$_{2}$) bond-clusters in graphene.
  }
  \begin{ruledtabular}
  \begin{tabular}{cccccc}
  C & $\bm{R}_{i}$ & B$_{1}$ & $\bm{b}_{i}@\bm{c}_{i}$ & B$_{2}$ & $\bm{b}_{i}@\bm{c}_{i}$ \\ \hline
  C$_{1}$ & $\left[\frac{2}{3},\frac{1}{3},0\right]$ & b$_{1}$ & $\left[\frac{2}{3},\frac{1}{3},0\right]@\left[[0,\frac{1}{2},0\right]$                        & b$_{4}$ & $\left[1, 0, 0\right]@\left[\frac{1}{6}, \frac{1}{3}, 0\right]$ \\
  C$_{2}$ & $\left[\frac{1}{3},\frac{2}{3},0\right]$  & b$_{2}$ & $\left[-\frac{1}{3}, \frac{1}{3}, 0\right]@\left[[\frac{1}{2}, \frac{1}{2}, 0\right]$     & b$_{5}$ & $\left[1, 0, 0\right]@\left[\frac{5}{6}, \frac{2}{3}, 0\right]$ \\
  &      & b$_{3}$ & $\left[-\frac{1}{3}, -\frac{2}{3}, 0\right]@\left[[\frac{1}{2}, 0, 0\right]$                  & b$_{6}$ & $\left[1, 1, 0\right]@\left[\frac{5}{6}, \frac{1}{6}, 0\right]$ \\
  &      &  &  & b$_{7}$ & $\left[0, 1, 0\right]@\left[\frac{1}{3}, \frac{1}{6}, 0\right]$ \\
  &      &  &  & b$_{8}$ & $\left[1, 1, 0\right]@\left[\frac{1}{6}, \frac{5}{6}, 0\right]$ \\
  &      &  &  & b$_{9}$ & $\left[0, 1, 0\right]@\left[\frac{2}{3}, \frac{5}{6}, 0\right]$ \\
  \end{tabular}
\end{ruledtabular}
\end{table}

\begin{table}[t]
  \caption{\label{tbl_graphene_sb_samb}
  SAMB in C site-cluster and B$_{1}$ and B$_{2}$ bond-clusters.
  }
  \begin{ruledtabular}
  \begin{tabular}{cccccc}
  Cluster & $l$ & $\Gamma$ & $\gamma$ & Symbol & Definition \\ \hline
  C          &  0   & $A_{1g}$   &  -                 &  $\mathbb{Q}_{0}^{\rm (C)}$    & $\frac{1}{\sqrt{2}}(1,1)$  \\
               &  3   & $B_{1u}$   & -                  &  $\mathbb{Q}_{3a}^{\rm (C)}$  & $\frac{1}{\sqrt{2}}(1,-1)$ \\
  \hline\hline
  B$_{1}$ & 0 & $A_{1g}$     & -                 & $\mathbb{Q}_{0}^{\rm (B_{1})}$   & $\frac{1}{\sqrt{3}}(1,1,1)$ \\
                & 2 & $E_{2g}$     & $u$            & $\mathbb{Q}_{v}^{\rm (B_{1})}$   & $\frac{1}{\sqrt{6}}(1,1,-2)$ \\
                &    &                     & $v$            & $\mathbb{Q}_{xy}^{\rm (B_{1})}$ & $\frac{1}{\sqrt{2}}(-1,1,0)$ \\
  \hline
          & 1 & $E_{1u}$     & $u$            & $\mathbb{T}_{x}^{\rm (B_{1})}$   & $\frac{i}{\sqrt{2}}(1,-1,0)$  \\
                &     &                    & $v$            & $\mathbb{T}_{y}^{\rm (B_{1})}$   & $\frac{i}{\sqrt{6}}(1,1,-2)$ \\
                & 3 & $B_{1u}$     & -                 & $\mathbb{T}_{3a}^{\rm (B_{1})}$ & $\frac{i}{\sqrt{3}}(1,1,1)$ \\
  \hline\hline
  B$_{2}$ & 0 & $A_{1g}$     & -                 & $\mathbb{Q}_{0}^{\rm (B_{2})}$   & $\frac{1}{\sqrt{6}}(1,1,1,1,1,1)$ \\
                & 1 & $E_{1u}$     & $u$            & $\mathbb{Q}_{x}^{\rm (B_{2})}$   & $\frac{1}{2}(0,0,-1,1,1,-1)$ \\
                &    &                     & $v$            & $\mathbb{Q}_{y}^{\rm (B_{2})}$   & $\frac{1}{2\sqrt{3}}(2,-2,1,1,-1,-1)$ \\
                & 2 & $E_{2g}$     & $u$            & $\mathbb{Q}_{v}^{\rm (B_{2})}$   & $\frac{1}{2\sqrt{3}}(2,2,-1,-1,-1,-1)$ \\
                &    &                     & $v$            & $\mathbb{Q}_{xy}^{\rm (B_{2})}$ & $\frac{1}{2}(0,0,-1,1,-1,1)$ \\
                & 3 & $B_{1u}$     & -                 & $\mathbb{Q}_{3a}^{\rm (B_{2})}$ & $\frac{1}{\sqrt{6}}(1,-1,-1,-1,1,1)$ \\
  \hline
          & 1 & $A_{2g}$     & -                 & $\mathbb{M}_{z}^{\rm (B_{2})}$ & $\frac{i}{\sqrt{6}}(1,-1,1,-1,-1,1)$ \\
          &    & $E_{1u}$     & $u$            & $\mathbb{T}_{x}^{\rm (B_{2})}$   & $\frac{i}{2\sqrt{3}}(2,2,1,-1,1,-1)$ \\
                &    &                     & $v$            & $\mathbb{T}_{y}^{\rm (B_{2})}$   & $\frac{i}{2}(0,0,1,1,1,1)$ \\
                & 2 & $E_{2g}$     & $u$            & $\mathbb{T}_{v}^{\rm (B_{2})}$   & $\frac{i}{2}(0,0,-1,-1,1,1)$ \\
                &    &                     & $v$            & $\mathbb{T}_{xy}^{\rm (B_{2})}$ & $\frac{i}{2\sqrt{3}}(-2,2,1,-1,-1,1)$ \\
                & 3 & $B_{2u}$     & -                 & $\mathbb{T}_{3a}^{\rm (B_{2})}$ & $\frac{i}{\sqrt{6}}(1,1,-1,1,-1,1)$ \\
  \end{tabular}
  \end{ruledtabular}
\end{table}

\begin{figure*}[t]
  \begin{center}
  \includegraphics[width=0.98 \hsize]{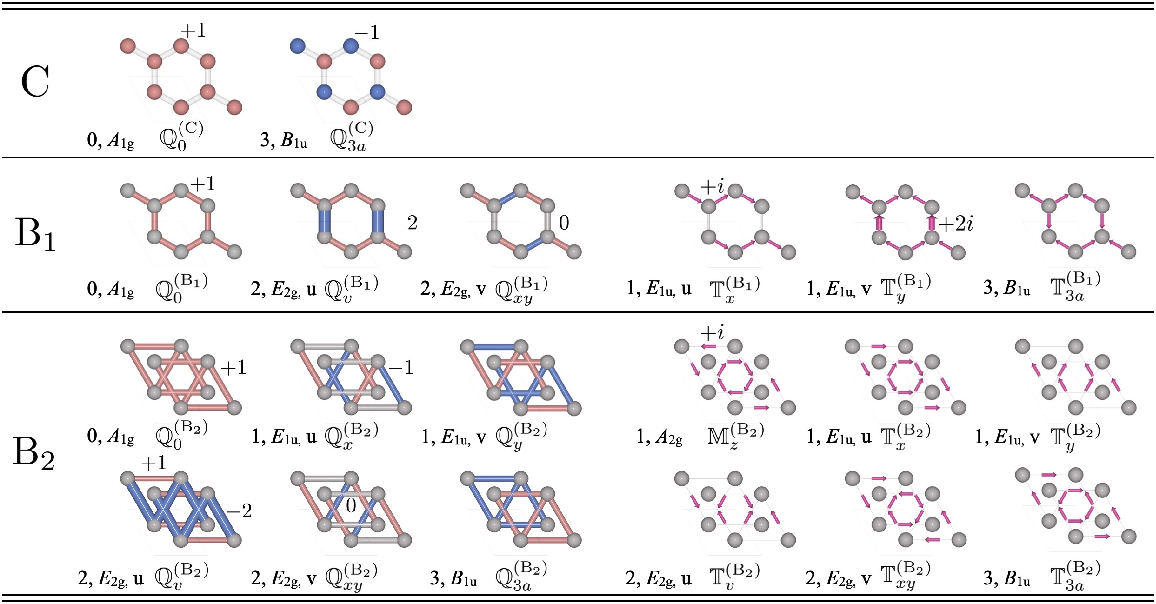}
  \vspace{5mm}
  \caption{
  \label{fig_sb_samb}
Schematic of the site-cluster SAMBs, and the nearest-neighbor and 2nd-neighbor bond-cluster SAMBs for monolayer graphene.
The color and size (width) represent the weight of the components in each SAMB, and the number written before the irrep. stands for the rank of SAMB.
  }
  \end{center}
\end{figure*}

In this section, we show the explicit definition of the cluster SAMB by taking a graphene as an example.
A honeycomb structure of monolayer graphene belongs to the space group $P6/mmm$ ($\#$191, D$_{\rm 6h}^{1}$) and its associated point group is D$_{\rm 6h}$.
As shown in Fig.~\ref{fig_sb_cluster}, there are two C atoms C$_{1}$ and C$_{2}$ within the unit cell.
The site-cluster C: (C$_{1}$, C$_{2}$), the bond clusters B$_{1}$: (b$_{1}$,b$_{2}$,b$_{3}$) and B$_{2}$: (b$_{4}$,b$_{5}$,\ldots,b$_{9}$) for three nearest-neighbor and six 2nd-neighbor bonds are summarized in Table~\ref{tbl_graphene_sb}.
Each bond is defined by b$_{i} = \bm{b}_{i}@\bm{c}_{i}$, where we have introduced a bond-vector $\bm{b}_{i} = \bm{R}_{\rm head} - \bm{R}_{\rm tail}$ and bond-center $\bm{c}_{i} = (\bm{R}_{\rm head} + \bm{R}_{\rm tail})/2$.

The site-cluster SAMB is defined by a real $N_{\rm s}$-dimensional vector basis ($N_{\rm s}$ is the number of sites in each site-cluster) by evaluating $O_{l\xi} (\bm{r})$ at each atomic site $\bm{r} = \bm{R}_{i}$ in the site cluster:
\begin{align}
\mathbb{Q}_{l\xi}^{\rm (s)} = (q_{1}^{(l\xi)},q_{2}^{(l\xi)},\cdots,q_{N_{\rm s}}^{(l\xi)}), \quad q_{i}^{(l\xi)} = O_{l\xi} (\bm{R}_{i}).
\label{eq_Qs_samb}
\end{align}
Hereafter, the superscript ``(s)'' indicates the site-cluster SAMB.
Note that the site-cluster SAMBs are defined as the E multipoles ``$\mathbb{Q}$'', and they describe the onsite degrees of freedom.

Similarly, the E bond-cluster SAMB is defined by a real $N_{\rm b}$-dimensional vector basis ($N_{\rm b}$ is the number of bonds in each bond-cluster) by evaluating $O_{l\xi} (\bm{r})$ at each bond center $\bm{r} = \bm{c}_{i}$ in each bond-cluster as
\begin{align}
\mathbb{Q}_{l\xi}^{\rm (b)}=(c_{1}^{(l\xi)},c_{2}^{(l\xi)},\cdots,c_{N_{b}}^{(l\xi)}), \quad c_{i}^{(l\xi)} = O_{l\xi} (\bm{c}_{i}).
\label{eq_Qb_samb}
\end{align}
Hereafter, the superscript ``(b)'' indicates the bond-cluster SAMB.
As shown in Table~\ref{tab_samb_hamiltonian}, $\mathbb{Q}_{l\xi}^{\rm (b)}$ can express the real hopping, $(c_{i}^{\dagger}c_{j}^{} + c_{j}^{\dagger}c_{i}^{})$.
Since the real hopping is symmetric for the reversal of the bond direction, the directional property of the bonds can be omitted.
Meanwhile, when we consider the antisymmetric bond dependence such as an imaginary hopping, e.g., $i(c_{i}^{\dagger}c_{j}^{} - c_{j}^{\dagger}c_{i}^{})$, the directional property must be taken into account.
Then, the bond-cluster SAMB for describing the antisymmetric bond dependence is defined by a complex $N_{\rm b}$-dimensional vector basis as
\begin{align}
\mathbb{T}_{l\xi}^{\rm (b)} = i(b_{1}^{(l\xi)},b_{2}^{(l\xi)},\cdots,b_{N_{b}}^{(l\xi)}), \quad b_{i}^{(l\xi)} = {\rm sgn}(\bm{b}_{i})|O_{l\xi} (\bm{c}_{i})|,
\label{eq_Tb_samb}
\end{align}
where ${\rm sgn}(\bm{b}_{i})$ is determined by the definition of the bond-cluster and its detailed explanation is given in Ref.~\cite{PhysRevB.98.165110}.
Here, we have attached the phase factor $i$ for describing the antisymmetric bond dependence.
``$\mathbb{T}$'' denotes the MT multipole, and it has a magnetic polar tensor property as well as the electric current.

Hereafter, the symbol $\mathbb{Y}_{l\xi}^{\rm (s/b)}$ is used to refer to the site-cluster SAMB $\mathbb{Q}_{l\xi}^{\rm (s)}$ and the bond-cluster SAMB $\mathbb{Q}_{l\xi}^{\rm (b)}$/$\mathbb{T}_{l\xi}^{\rm (b)}$.
The site/bond-cluster SAMBs are transformed by the space group symmetry operation $R^{\rm (s)}(g)$ given by Eq.~(\ref{eq_sym_oper_AO_R}) as
\begin{align}
  R^{\rm (s)}(g) \mathbb{Y}_{l \Gamma n \gamma}^{\rm (s/b)} R^{\rm (s) -1}(g) = \sum_{\gamma'} \mathbb{Y}_{l \Gamma n \gamma'}^{\rm (s/b)} D_{\gamma',\gamma}^{\Gamma} (p),
\end{align}
where $D^{\Gamma} (p)$ is the representation matrix of $\Gamma$ irrep.
The site-cluster SAMB $\mathbb{Q}_{l \xi}^{\rm (s)}$ satisfies the orthonormal and complete relations in each site-cluster as
\begin{align}
& \sum_{l,\xi} \mathbb{Q}_{l \xi}^{\rm (s)}(\bm{R}_{i}) \mathbb{Q}_{l\xi}^{\rm (s)}(\bm{R}_{j}) = \delta_{i,j},
\label{eq_s_samb_comp}
\\ &
\mathrm{Tr} \left[\mathbb{Q}_{l \xi}^{\rm (s)} \mathbb{Q}_{l' \xi'}^{\rm (s)} \right] = \delta_{l,l'} \delta_{\xi,\xi'}.
\label{eq_s_samb_ortho}
\end{align}
Similarly, the bond-cluster SAMB $\mathbb{Y}_{l \xi}^{\rm (b)}$ satisfies the orthonormal and complete relations in each bond-cluster as
\begin{align}
& \sum_{l,\xi} \mathbb{Y}_{l \xi}^{\rm (b)}(\bm{b}_{i}@\bm{c}_{i}) \mathbb{Y}_{l\xi}^{\rm (b)}(\bm{b}_{j}@\bm{c}_{j}) = \delta_{i,j},
\label{eq_b_samb_comp}
\\ &
\mathrm{Tr} \left[\mathbb{Y}_{l \xi}^{\rm (b)} \mathbb{Y}_{l' \xi'}^{\rm (b)'} \right] = \delta_{\mathbb{Y},\mathbb{Y}'} \delta_{l,l'} \delta_{\xi,\xi'}.
\label{eq_b_samb_ortho}
\end{align}

Following Eqs.~(\ref{eq_Qs_samb}), (\ref{eq_Qb_samb}), and (\ref{eq_Tb_samb}), the site/bond-cluster SAMBs in the graphene are obtained as summarized in Table~\ref{tbl_graphene_sb_samb}, and they are schematically shown in Fig.~\ref{fig_sb_samb}.

\subsection{Atomic SAMB}
\label{sec_atomic_samb}

\begin{table*}[t]
  \renewcommand{\arraystretch}{1.1}
  \begin{center}
  \caption{ \label{tab_a_samb}
  Operator and matrix expressions of the orthonormalized atomic SAMB up to rank 2 in the point group D$_{\rm 6h}$.
  We consider the spinful $(s; p_{x}, p_{y}, p_{z})$ orbitals in each C site, and $\braket{s|s}$, $\braket{s|p}$, and $\braket{p|p}$ Hilbert spaces.
  The upper and lower parts separated by the double line represent the spinless and spinful atomic SAMBs, respectively.
  We omit $(s,k)=(0,0)$ for the spinless SAMBs for simplicity, because only $s=k=0$ (charge sector) SAMBs are active in the spinless Hilbert space.
  $\bm{l}$ and $\bm{\sigma}/2$ represent the dimensionless orbital and spin angular-momentum operators.
  We have introduced the abbreviations, $Q_{u}^{(\pm)} = (Q_{u}^{\rm (a)} \pm \sqrt{3}Q_{v}^{\rm (a)})/2$, $Q_{v}^{(\pm)} = (\pm \sqrt{3}Q_{u}^{\rm (a)} + Q_{v}^{\rm (a)}) / 2$, for notational simplicity.
  }
  \scalebox{0.96}[0.96]{
  \begin{tabular}{ccccccccccc}
  \hline\hline
  $\mathcal{H}$  & $l$ & $\Gamma$ & $\gamma$ & Symbol & Definition & $l$ & $\Gamma$ & $\gamma$ & Symbol & Definition
  \\
  \hline
  $\braket{s|s}$ &  0  & $A_{1g}$ & $-$ & $\mathbb{Q}_{0, s}^{\rm (a)}$ & $1, \frac{1}{\sqrt{2}}\begin{pmatrix}1\end{pmatrix}\otimes\sigma_{0}$ \\
  $\braket{s|p}$ &  1  & $A_{2u}$ & $-$ & $\mathbb{Q}_{z}^{\rm (a)}$ & $z, \frac{1}{\sqrt{2}}\begin{pmatrix}0 & 0 & 1\end{pmatrix}\otimes\sigma_{0}$  & 1  & $E_{1u}$ & $u,v$ & $\mathbb{Q}_{x}^{\rm (a)}, \mathbb{Q}_{y}^{\rm (a)}$ & $[x,y],  \frac{1}{\sqrt{2}} \left[\begin{pmatrix}1 & 0 & 0\end{pmatrix}\otimes\sigma_{0}, \begin{pmatrix}0 & 1 & 0\end{pmatrix}\otimes\sigma_{0}\right]$ \\
                             &    & $A_{2u}$ & $-$ & $\mathbb{T}_{z}^{\rm (a)}$ & $(\bm{r}\times\bm{l})_{z}, \frac{1}{\sqrt{2}}\begin{pmatrix}0 & 0 & i\end{pmatrix}\otimes\sigma_{0}$ &   & $E_{1u}$ & $u,v$ & $\mathbb{T}_{x}^{\rm (a)}, \mathbb{T}_{y}^{\rm (a)}$ & $[(\bm{r}\times\bm{l})_{x}, (\bm{r}\times\bm{l})_{y}], \frac{1}{\sqrt{2}} \left[\begin{pmatrix}i & 0 & 0\end{pmatrix}\otimes\sigma_{0}, \begin{pmatrix}0 & i & 0\end{pmatrix}\otimes\sigma_{0}\right]$\\
  $\braket{p|p}$ &  0  & $A_{1g}$ & $-$ & $\mathbb{Q}_{0, p}^{\rm (a)}$ & $1, \frac{1}{\sqrt{6}}\begin{pmatrix}1&0&0 \\ 0&1&0 \\ 0&0&1\end{pmatrix}\otimes\sigma_{0}$ \\
                             &  1  & $A_{2g}$ & $-$ & $\mathbb{M}_{z}^{\rm (a)}$ & $l_{z}, \frac{1}{2}\begin{pmatrix}0 & -i & 0 \\ i & 0 & 0 \\ 0 & 0 & 0 \end{pmatrix}\otimes\sigma_{0}$ & 1  & $E_{1g}$ & $u,v$ & $-\mathbb{M}_{y}^{\rm (a)}, \mathbb{M}_{x}^{\rm (a)}$ & $[-l_{y}, l_{x}], \frac{1}{2} \left[\begin{pmatrix}0 & 0 & -i \\ 0 & 0 & 0 \\ i & 0 & 0 \end{pmatrix}\otimes\sigma_{0}, \begin{pmatrix}0 & 0 & 0 \\ 0 & 0 & -i \\ 0 & i & 0 \end{pmatrix}\otimes\sigma_{0}\right]$\\
                       &  2  & $A_{1g}$ & $-$ & $\mathbb{Q}_{u}^{\rm (a)}$ & $3z^{2} - r^{2}, \frac{1}{2\sqrt{3}}\begin{pmatrix}-1&0&0 \\ 0&-1&0 \\ 0&0&2\end{pmatrix}\otimes\sigma_{0}$   & 2  & $E_{1g}$ & $u,v$ & $\mathbb{Q}_{zx}^{\rm (a)},\mathbb{Q}_{yz}^{\rm (a)}$ & $[zx, yz], \frac{1}{2} \left[ \begin{pmatrix}0&0&1 \\ 0&0&0 \\ 1&0&0\end{pmatrix}\otimes\sigma_{0}, \begin{pmatrix}0&0&0 \\ 0&0&1 \\ 0&1&0\end{pmatrix}\otimes\sigma_{0} \right]$   \\
      &&&&&&              &    $E_{2g}$ & $u, v$ & $\mathbb{Q}_{v}^{\rm (a)}, -\mathbb{Q}_{xy}^{\rm (a)}$  & $[x^{2}-y^{2}, -xy], \frac{1}{2} \left[ \begin{pmatrix}1&0&0 \\ 0&-1&0 \\ 0&0&0\end{pmatrix}\otimes\sigma_{0}, - \begin{pmatrix}0&1&0 \\ 1&0&0 \\ 0&0&0\end{pmatrix}\otimes\sigma_{0} \right]$\\
  \hline\hline
  $\braket{s|s}$ &  1  & $A_{2g}$ & $-$ & $\mathbb{M}_{z, s, (1,-1)}^{\rm (a)}$ & $\mathbb{Q}_{0, s}^{\rm (a)} \sigma_{z}$   &  1  & $E_{1g}$ & $u,v$ & $-\mathbb{M}_{y, s, (1,-1)}^{\rm (a)}, \mathbb{M}_{x, s, (1,-1)}^{\rm (a)}$ & $\mathbb{Q}_{0, s}^{\rm (a)} \left[ -\sigma_{y},  \sigma_{x} \right]$ \\
  $\braket{s|p}$ &  0  & $A_{1u}$ & $-$ & $\mathbb{G}_{0, (1,1)}^{\rm (a)}$ & $\frac{1}{\sqrt{3}} \bm{T}^{\rm (a)} \cdot \bm{\sigma}$  &  0  & $A_{1u}$ & $-$ & $\mathbb{M}_{0, (1,1)}^{\rm (a)}$ & $\frac{1}{\sqrt{3}} \bm{Q}^{\rm (a)} \cdot \bm{\sigma}$ \\
                         &  1  & $A_{2u}$ & $-$ & $\mathbb{Q}_{z, (1,0)}^{\rm (a)}$ & $\frac{1}{\sqrt{2}} (\bm{\sigma} \times  \bm{T}^{\rm (a)})_{z}$  &  1  & $E_{1u}$ & $u,v$ & $\mathbb{Q}_{x,(1,0)}^{\rm (a)}, \mathbb{Q}_{y,(1,0)}^{\rm (a)}$ & $ \frac{1}{\sqrt{2}} \left[(\bm{\sigma} \times \bm{T}^{\rm (a)})_{x}, (\bm{\sigma} \times \bm{T}^{\rm (a)})_{y}\right]$ \\
                         &      & $A_{2u}$ & $-$ & $\mathbb{T}_{z,(1,0)}^{\rm (a)}$ & $\frac{1}{\sqrt{2}} (\bm{\sigma} \times \bm{Q}^{\rm (a)})_{z}$   &    & $E_{1u}$ & $u,v$ & $\mathbb{T}_{x,(1,0)}^{\rm (a)}, \mathbb{T}_{y,(1,0)}^{\rm (a)}$ & $ \frac{1}{\sqrt{2}} \left[(\bm{\sigma} \times \bm{Q}^{\rm (a)})_{x}, (\bm{\sigma} \times \bm{Q}^{\rm (a)})_{y}\right]$\\
                         &  2  & $A_{1u}$ & $-$ & $\mathbb{G}_{u,(1,-1)}^{\rm (a)}$ & $ \frac{1}{\sqrt{6}} \left(2 \sigma_{z} \mathbb{T}_{z}^{\rm (a)} - \sigma_{x} \mathbb{T}_{x}^{\rm (a)} -\sigma_{y} \mathbb{T}_{y}^{\rm (a)} \right)$                       &  2  & $E_{1u}$ & $u,v$ & $-\mathbb{G}_{yz, (1,-1)}^{\rm (a)},\mathbb{G}_{zx, (1,-1)}^{\rm (a)}$ & $\frac{1}{\sqrt{2}} \left[ -(\sigma_{y} \mathbb{T}_{z}^{\rm (a)} + \sigma_{z} \mathbb{T}_{y}^{\rm (a)} ), (\sigma_{z} \mathbb{T}_{x}^{\rm (a)} + \sigma_{x} \mathbb{T}_{z}^{\rm (a)} ) \right]$  \\
      &&&&&&              &    $E_{2u}$ & $u,v$ & $\mathbb{G}_{xy, (1,-1)}^{\rm (a)}, \mathbb{G}_{v, (1,-1)}^{\rm (a)}$ & $\frac{1}{\sqrt{2}} \left[ (\sigma_{x} \mathbb{T}_{y}^{\rm (a)} + \sigma_{y}  \mathbb{T}_{x}^{\rm (a)} ), (\sigma_{x} \mathbb{T}_{x}^{\rm (a)} - \sigma_{y} \mathbb{T}_{y}^{\rm (a)} ) \right]$  \\
                                                  &  &  $A_{1u}$ & $-$ & $\mathbb{M}_{u, (1,-1)}^{\rm (a)}$ & $ \frac{1}{\sqrt{6}} \left(2 \sigma_{z} \mathbb{Q}_{z}^{\rm (a)} - \sigma_{x} \mathbb{Q}_{x}^{\rm (a)} - \sigma_{y} \mathbb{Q}_{y}^{\rm (a)} \right)$  &   & $E_{1u}$ & $u,v$ & $-\mathbb{M}_{yz, (1,-1)}^{\rm (a)},\mathbb{M}_{zx, (1,-1)}^{\rm (a)}$ & $\frac{1}{\sqrt{2}} \left[ -(\sigma_{y} \mathbb{Q}_{z}^{\rm (a)} + \sigma_{z} \mathbb{Q}_{y}^{\rm (a)} ), (\sigma_{z} \mathbb{Q}_{x}^{\rm (a)} + \sigma_{x} \mathbb{Q}_{z}^{\rm (a)} ) \right]$ \\
      &&&&&&              &    $E_{2u}$ & $u,v$ & $\mathbb{M}_{xy, (1,-1)}^{\rm (a)},\mathbb{M}_{v, (1,-1)}^{\rm (a)}$ & $\frac{1}{\sqrt{2}} \left[ (\sigma_{x} \mathbb{Q}_{y}^{\rm (a)} + \sigma_{y}  \mathbb{Q}_{x}^{\rm (a)} ), (\sigma_{x} \mathbb{Q}_{x}^{\rm (a)} - \sigma_{y} \mathbb{Q}_{y}^{\rm (a)} ) \right] $ \\
  $\braket{p|p}$ &  0  & $A_{1g}$ & $-$ & $\mathbb{Q}_{0, (1,1)}^{\rm (a)}$ & $\frac{1}{\sqrt{3}} \bm{M}^{\rm (a)}\cdot\bm{\sigma} $  \\
                      &  1   & $A_{2g}$ & $-$ & $\mathbb{G}_{z, (1,0)}^{\rm (a)}$ & $\frac{1}{\sqrt{2}} (\bm{\sigma} \times \bm{M}^{\rm (a)})_{z}$   &  1  & $E_{1g}$ & $u,v$ & $-\mathbb{G}_{y, (1,0)}^{\rm (a)}, \mathbb{G}_{x, (1,0)}^{\rm (a)}$ & $ \frac{1}{\sqrt{2}} \left[-(\bm{\sigma} \times \bm{M}^{\rm (a)})_{y}, (\bm{\sigma} \times \bm{M}^{\rm (a)})_{x}\right]$\\
                      &     & $A_{2g}$ & $-$ & $\mathbb{M}_{z, p, (1,-1)}^{\rm (a)}$ & $\mathbb{Q}_{0, p}^{\rm (a)} \sigma_{z}$   &     & $E_{1g}$ & $u,v$ & $-\mathbb{M}_{y, p, (1,-1)}^{\rm (a)}, \mathbb{M}_{x, p, (1,-1)}^{\rm (a)}$ & $\mathbb{Q}_{0, p}^{\rm (a)}  \left[-\sigma_{y},  \sigma_{x} \right]$ \\
                      &     & $A_{2g}$ & $-$ & $\mathbb{M}_{z, a, (1,1)}^{\rm (a)}$ & $ \sqrt{\frac{3}{10}} \left(\sigma_{x} \mathbb{Q}_{zx}^{\rm (a)} + \sigma_{y} \mathbb{Q}_{yz}^{\rm (a)} +  \frac{2}{\sqrt{3}} \sigma_{z}  \mathbb{Q}_{u}^{\rm (a)} \right)$   &     & $E_{1g}$ & $\left[ \begin{matrix} u \\ v \end{matrix} \right]$ & $\left[ \begin{matrix} -\mathbb{M}_{y, a, (1,1)}^{\rm (a)} \\ \mathbb{M}_{x, a, (1,1)}^{\rm (a)} \end{matrix} \right]$ & $ \sqrt{\frac{3}{10}} \left[ \begin{matrix} \left( - \sigma_{x} \mathbb{Q}_{xy}^{\rm (a)} -  \frac{1}{\sqrt{3}} \sigma_{y} \mathbb{Q}_{u}^{(+)} + \sigma_{z} \mathbb{Q}_{yz}^{\rm (a)} \right) \\ \left( -  \frac{1}{\sqrt{3}}  \sigma_{x} \mathbb{Q}_{u}^{(-)} - \sigma_{y} \mathbb{Q}_{xy}^{\rm (a)} + \sigma_{z} \mathbb{Q}_{zx}^{\rm (a)} \right) \end{matrix} \right] $ \\
                         &  2  & $A_{1g}$ & $-$ & $\mathbb{Q}_{u, (1,-1)}^{\rm (a)}$ & $ \frac{1}{\sqrt{6}} \left(2 \sigma_{z} \mathbb{M}_{z}^{\rm (a)} - \sigma_{x} \mathbb{M}_{x}^{\rm (a)} -\sigma_{y} \mathbb{M}_{y}^{\rm (a)} \right)$                       &  2  & $E_{1g}$ & $u,v$ & $\mathbb{Q}_{zx, (1,-1)}^{\rm (a)},\mathbb{Q}_{yz, (1,-1)}^{\rm (a)}$ & $\frac{1}{\sqrt{2}} \left[ (\sigma_{z} \mathbb{M}_{x}^{\rm (a)} + \sigma_{x} \mathbb{M}_{z}^{\rm (a)} ), -(\sigma_{y} \mathbb{M}_{z}^{\rm (a)} + \sigma_{z} \mathbb{M}_{y}^{\rm (a)} ) \right]$  \\
      &&&&&&              &    $E_{2g}$ & $u,v$ & $\mathbb{Q}_{v, (1,-1)}^{\rm (a)}, -\mathbb{Q}_{xy, (1,-1)}^{\rm (a)}$ & $\frac{1}{\sqrt{2}} \left[ (\sigma_{x} \mathbb{M}_{x}^{\rm (a)} - \sigma_{y} \mathbb{M}_{y}^{\rm (a)} ), -(\sigma_{x} \mathbb{M}_{y}^{\rm (a)} + \sigma_{y}  \mathbb{M}_{x}^{\rm (a)} )\right]$  \\
                         &    & $A_{1g}$ & $-$ & $\mathbb{T}_{u, (1,0)}^{\rm (a)}$ & $ \frac{1}{\sqrt{2}} \left(\sigma_{x} \mathbb{Q}_{yz}^{\rm (a)} - \sigma_{y} \mathbb{Q}_{zx}^{\rm (a)} \right)$                       &

                         & $E_{1g}$ & $\left[ \begin{matrix} u \\ v \end{matrix} \right]$ & $\left[ \begin{matrix}  \mathbb{T}_{zx, (1,0)}^{\rm (a)} \\ \mathbb{T}_{yz, (1,0)}^{\rm (a)} \end{matrix} \right] $ & $\frac{1}{\sqrt{6}} \left[ \begin{matrix} ( -\sigma_{x} \mathbb{Q}_{xy}^{\rm (a)} - \sigma_{y} \mathbb{Q}_{v}^{(-)} - \sigma_{z} \mathbb{Q}_{yz}^{\rm (a)}) \\ (-\sigma_{x} \mathbb{Q}_{v}^{(+)} + \sigma_{y} \mathbb{Q}_{xy}^{\rm (a)} + \sigma_{z} \mathbb{Q}_{zx}^{\rm (a)}) \end{matrix} \right]$  \\
      &&&&&&             & $E_{2g}$ & $\left[ \begin{matrix} u \\ v \end{matrix} \right]$ & $\left[ \begin{matrix}  \mathbb{T}_{v, (1,0)}^{\rm (a)} \\ -\mathbb{T}_{xy, (1,0)}^{\rm (a)} \end{matrix} \right] $ & $\frac{1}{\sqrt{6}} \left[ \begin{matrix} ( \sigma_{x} \mathbb{Q}_{yz}^{\rm (a)} + \sigma_{y} \mathbb{Q}_{zx}^{\rm (a)} + 2 \sigma_{z} \mathbb{Q}_{xy}^{\rm (a)}) \\ (-\sigma_{x} \mathbb{Q}_{zx}^{\rm (a)} + \sigma_{y} \mathbb{Q}_{yz}^{\rm (a)} + 2 \sigma_{z} \mathbb{Q}_{v}^{\rm (a)}) \end{matrix} \right]$  \\
  \hline\hline
  \end{tabular}
  }
  \end{center}
\end{table*}

The spinful atomic SAMB including atomic orbital and spin degrees of freedom is defined by the direct product of the spinless atomic SAMB $\mathbb{X}_{lm}^{\rm (orb)}$ and identity or Pauli matrices ($\sigma_{00} = \sigma_{0}, \sigma_{10} = \sigma_{z}, \sigma_{1\pm 1} = \mp (\sigma_{x}\pm i \sigma_{y})/\sqrt{2}$) as
\begin{align}
\mathbb{X}_{lm, (s,k)}^{\rm (a)}=i^{s+k} \sum \braket{l+k, m-n ; s n | l m} \mathbb{X}_{l+k, m-n}^{\rm (orb)} \sigma_{sn}^{},
\end{align}
where $s = 0, 1$ and $k = -s, 0, s$.
Here, $\braket{l+k, m-n ; s n | l m}$ is the Clebsch-Gordan (CG) coefficient, which is introduced for the addition rule of the angular momentum.
Note that $\mathbb{X}_{lm, (0,0)}^{\rm (a)} = \mathbb{X}_{lm}^{\rm (orb)}$ is the spinless SAMB, while $\mathbb{X}_{lm, (s,k)}^{\rm (a)}$ ($s = 1, k = 0, \pm 1$) is the spinful atomic SAMB.
Hereafter, the superscript ``(a)'' indicates the atomic SAMB and the symbol $\mathbb{X}_{lm}$ is used to refer to all of the four-type atomic SAMBs.
See Ref.~\cite{PhysRevB.98.165110} for detailed definition of $\mathbb{X}_{lm}^{\rm (orb)}$ and the formula to compute the matrix elements of $\mathbb{X}_{lm, (s,k)}^{\rm (a)}$ is given in Ref.~\cite{JPSJ.89.104704}.

Using the unitary matrix given by Eq.~(\ref{eq_Olxi}), the symmetry-adapted atomic SAMB for point group is obtained by
\begin{align}
\mathbb{X}_{l \xi, (s,k)}^{\rm (a)} = \sum_{m} U_{m, \xi}^{(l)} \mathbb{X}_{lm, (s,k)}^{\rm (a)}.
\end{align}
$\mathbb{X}_{l \xi, (s,k)}^{\rm (a)}$ is transformed by the point group symmetry operation $R^{\rm (a)}(p)$ given by Eq.~(\ref{eq_sym_oper_AO}) as
\begin{align}
  R^{\rm (a)}(p) \mathbb{X}_{l \Gamma n \gamma, (s, k)}^{\rm (a)} R^{\rm (a) -1}(p) = \sum_{\gamma'} \mathbb{X}_{l \Gamma n \gamma', (s,k)}^{\rm (a)} D_{\gamma',\gamma}^{\Gamma} (p),
\end{align}
where $D^{\Gamma} (p)$ is the representation matrix of $\Gamma$ irrep.
Then, we normalize $\mathbb{X}_{l \xi, (s,k)}^{\rm (a)}$ as satisfying the orthonormal and complete relations:
\begin{align}
& \sum_{\mathbb{X},l,\xi,s,k} \left[\mathbb{X}_{l \xi, (s,k)}^{\rm (a)}\right]_{a_{1}a_{2}}  \left[\mathbb{X}_{l\xi, (s,k)}^{\rm (a)}\right]_{a_{3}a_{4}} = \delta_{a_{1},a_{4}} \delta_{a_{2},a_{3}},
\label{eq_a_samb_comp}
\\ &
\mathrm{Tr} \left[\mathbb{X}_{l \xi, (s,k)}^{\rm (a)} \mathbb{X}_{l' \xi', (s',k')}^{\rm (a)'} \right] = \delta_{\mathbb{X},\mathbb{X}'} \delta_{l,l'} \delta_{\xi,\xi'} \delta_{s,s'}\delta_{k,k'}.
\label{eq_a_samb_ortho}
\end{align}
The spinless and spinful atomic SAMBs up to rank 2 defined in the $s$, $p_{x}$, $p_{y}$, and $p_{z}$ orbitals are summarized in Table~\ref{tab_a_samb}.
Although the rank of the multipoles is no longer strictly preserved, it is partially inherited from their original definition.
Then, we can use this atomic SAMBs for the Hilbert space of the $s$, $p_{x}$, $p_{y}$, and $p_{z}$ like CWFs of graphene.
Any parity specific anisotropy of the electronic internal degrees of freedom at each C atom are described by the atomic SAMBs.

As summarized in Table~\ref{tab_a_samb}, there are three spinful monopoles.
The spinful E monopole $\mathbb{Q}_{0, (1,1)}^{\rm (a)}$ in $\braket{p|p}$ can describe the atomic SOC as shown in Table~\ref{tab_samb_hamiltonian}.
The spinful M monopole $\mathbb{M}_{0, (1,1)}^{\rm (a)}$ in $\braket{s|p}$ is the microscopic origin of the longitudinal magnetoelectric response~\cite{Spaldin_ME_2008}.
Recent studies have developed a systematic formalism to numerically evaluate the macroscopic M monopole based on DFT framework~\cite{PhysRevB.76.214404, PhysRevB.88.094429, PhysRevB.93.195167}.
On the other hand, the spinful ET monopole $\mathbb{G}_{0, (1,1)}^{\rm (a)}$ in $\braket{s|p}$ has the $\mathcal{T}$-even pseudoscalar property and corresponds to the chirality at the quantum-mechanical level~\cite{PhysRevB.98.165110,Kishine_IJC_2022, Inda_JCP_2024}.
Recently, the staggered alignment of the spinful ET monopole $\mathbb{G}_{0,(1,1)}^{\rm (a)}$ is considered as order-parameter candidate for the hidden order in URu$_{2}$Si$_{2}$~\cite{fujimotoSpinNematicState2011, ikedaEmergentRank5Nematic2012c, chandraHastaticOrderHeavyfermion2013, PhysRevB.97.235142, Kambe_SCES_2019, Hayami_URu2Si2_2023}.

Additionally, there are three spinless dipoles, $\mathbb{Q}_{\mu}^{\rm (a)}$ and $\mathbb{T}_{\mu}^{\rm (a)}$ in $\braket{s|p}$, and $\mathbb{M}_{\mu}^{\rm (a)}$ in $\braket{p|p}$, and six spinful dipoles, $\mathbb{M}_{\mu, s, (1,-1)}$ in $\braket{s|s}$, $\mathbb{Q}_{\mu, (1,0)}^{\rm (a)}$ and $\mathbb{T}_{\mu, (1,0)}^{\rm (a)}$ in $\braket{s|p}$, and $\mathbb{M}_{\mu, p, (1,-1)}$, $\mathbb{M}_{\mu, a, (1,-1)}$, and $\mathbb{G}_{\mu,(1,0)}$ in $\braket{p|p}$.
As summarized in Table~\ref{tab_samb_external_field_response}, the E, M, MT, and ET dipoles are coupled with the electric field $\bm{E}$ or thermal gradient $-\bm{\nabla} T$, magnetic field $\bm{H}$, electric (thermal) current $\bm{J} (\bm{J}^{Q})$, and rotational distortion $\bm{\omega}$, respectively.
Therefore, once the perpendicular electric field along $c$ axis is applied to the monolayer graphene and the symmetry of the system is lowered from D$_{\rm 6h}$ to C$_{\rm 6v}$, the spinless and spinful atomic dipoles $\mathbb{Q}_{z}^{\rm (a)}$ and $\mathbb{Q}_{z, (1,0)}^{\rm (a)}$ will appear in the Hamiltonian as discussed in Sec.~\ref{sec_graphene_E_field}.
The anisotropic M dipole $\mathbb{M}_{\mu, a, (1,-1)}$ defined in the spinful $\braket{p|p}$ space has been extensively studied since it can be observed in x-ray magnetic circular dichroism (XMCD) measurements~\cite{STOHR1995253, STOHR1999470, Yamasaki_JPSJ_2020, Sasabe_PRL_2023} and plays an important role in the anomalous Hall effect~\cite{SH_Ma_AHE_2021}.
The spinless MT dipole $\mathbb{T}_{\mu}^{\rm (a)}$ in $\braket{s|p}$ is induced by the parity-mixing hybridization and it plays an important role for stabilizing the $s$-$p$ hybridized orbitals in monolayer graphene as shown later.
Moreover, the MT dipoles $\mathbb{T}_{\mu}^{\rm (a)}$ and $\mathbb{T}_{\mu, (1,0)}^{\rm (a)}$ can be regarded as the atomic-scale origin of the off-diagonal magnetoelectric effect~\cite{MY_T_JPSJ_2019}.
The ET dipole $\mathbb{G}_{\mu,(1,0)}$ in $\braket{p|p}$ has attracted interest since it is considered as the atomic-scale order-parameter for the ferro-rotational order observed in RbFe(MoO$_{4}$)$_{2}$~\cite{Jin_RbFeMoO42_2020, Hayashida_RbFeMoO42_2021}, NiTiO$_{3}$~\cite{Hayashida_ferro_rotation_2021}, Ca$_{5}$Ir$_{3}$O$_{12}$~\cite{Hasegawa_ferro_rotation_2020, Hanate_ferro_rotation_2021, SH_ferro_rotation_2023}, CaMn$_{7}$O$_{12}$~\cite{PhysRevLett.108.067201}, BaCoSiO$_{4}$~\cite{PhysRevB.105.184407}, and so on.
Recent studies have developed a systematic formalism to numerically evaluate $\mathbb{G}_{\mu,(1,0)}$ based on DFT framework~\cite{Bhowal_ET_dipole_2024}.

In this way, using atomic SAMB allows us to unveil the atomic-scale multipole degrees of freedom hidden in the Hilbert space of the CWFs and to predict possible unconventional responses.

\subsection{Combined SAMB}
\label{sec_combined_samb}

In the previous subsections, we have separately constructed the complete and orthonormal atomic and cluster SAMBs.
Then, by performing the irreducible decomposition of the direct product of atomic and cluster SAMBs, we finally obtain the combined SAMB as
\begin{align}
  & \mathbb{Z}_{j}
  =
  \sum_{\xi_1 \xi_2} C_{l \xi}^{l_1 \xi_1, l_2 \xi_2}(X, Y \mid Z)\, \mathbb{X}_{l_1 \xi_1, (s,k)}^{\rm (a)} \otimes \mathbb{Y}_{l_2 \xi_2}^{\rm (s/b)}.
\label{eq_c_samb_R}
\end{align}
Here we have introduced an abbreviation $j = (Z, l, \Gamma, n, \gamma, s, k)$ where $Z = Q,M,T,G$.
$C_{l \xi}^{l_1 \xi_1, l_2 \xi_2}(X,Y|Z)$ is an orthonormal CG coefficient that arises from the reduction to the irreducible representation $\Gamma$ from the direct product of $\Gamma_{1}$ and $\Gamma_{2}$ (See Ref.~\cite{PhysRevB.98.165110} in more detail).
That matrix elements of the combined SAMB is given by
\begin{multline}
  \braket{w^{\rm CW}_{a_{1}\alpha_{1}\bm{R}_{1}}|\mathbb{Z}_{j}|w^{\rm CW}_{a_{2}\alpha_{2}\bm{R}_{2}}}
  \\
  = \sum_{\xi_1 \xi_2} C_{l \xi}^{l_1 \xi_1, l_2 \xi_2}(X, Y \mid Z) \left[\mathbb{X}_{l_1 \xi_1, (s k)}^{\rm (a)}\right]_{a_{1}a_{2}} \otimes \mathbb{Y}_{l_2 \xi_2}^{\rm (s/b)}(\bm{b}@\bm{c}),
\end{multline}
where we have introduced a bond-vector $\bm{b} = (\bm{R}_{1}+\bm{\alpha}_{1}) - (\bm{R}_{2}+\bm{\alpha}_{2})$ and bond-center $\bm{c} = [(\bm{R}_{1}+\bm{\alpha}_{1}) + (\bm{R}_{2}+\bm{\alpha}_{2})]/2$.

Since $\mathbb{X}_{l\xi, (s,k)}^{\rm (a)}$, $\mathbb{Y}_{l\xi}^{\rm (s/b)}$, and $C_{l \xi}^{l_1 \xi_1, l_2 \xi_2}(X,Y|Z)$ are already orthonormalized, the combined SAMB $\mathbb{Z}_{j}$ satisfies the orthonormal conditions and completeness as well:
\begin{align}
&
\sum_{j} \braket{w^{\rm CW}_{a_{1}\alpha_{1}\bm{R}_{1}}|\mathbb{Z}_{j}|w^{\rm CW}_{a_{2}\alpha_{2}\bm{R}_{2}}} \braket{w^{\rm CW}_{a_{3}\alpha_{3}\bm{R}_{3}}|\mathbb{Z}_{j}|w^{\rm CW}_{a_{4}\alpha_{4}\bm{R}_{4}}} = \delta_{1,4} \delta_{2,3},
\label{eq_c_samb_R_comp}
\\ &
\mathrm{Tr}\left[\mathbb{Z}_{i} \mathbb{Z}_{j}\right] = \delta_{i,j}.
\label{eq_c_samb_R_ortho}
\end{align}

The momentum-space representation of Eq.~(\ref{eq_c_samb_R}) is given by
\begin{align}
\mathbb{Z}_{j}(\bm{k}) = \sum_{\xi_{1}\xi_{2}} C_{l\xi}^{l_{1}\xi_{1},l_{2}\xi_{2}}(X,Y|Z)\,\mathbb{X}_{l_{1}\xi_{1},(s,k)}^{\rm (a)}\,\otimes\mathbb{Y}_{l_{2}\xi_{2}}^{\rm (s/b)}(\bm{k}),
\label{eq_c_samb_k}
\end{align}
where
\begin{align}
 [\mathbb{Y}_{l_{2}\xi_{2}}^{\rm (s/b)}(\bm{k})]_{\alpha_{1}\alpha_{2}} =
  \begin{cases}
    \delta_{\alpha_{1},\alpha_{2}}\mathbb{Y}_{l \xi}^{\rm (s)}(\bm{\alpha}_{1}), \\
    \frac{1}{\sqrt{2}}\sum_{\bm{b}}' \left( e^{-i\bm{k}\cdot\bm{b}} \mathbb{Y}_{l \xi}^{\rm (b)}(\bm{b}@\bm{c}) + {\rm H.c.}\right).
  \end{cases}
  \label{eq_sb_samb_k}
\end{align}
Note that the expression of the site-cluster SAMB does not change in the momentum space, and the summation $\sum_{\bm{b}}'$ in the second line of Eq.~(\ref{eq_sb_samb_k}) means that the summation is taken over the bonds $\bm{b} = \bm{\alpha}_{1} - \bm{\alpha}_{2} \in$ lattice vector.
The binary operator in Eq.~(\ref{eq_c_samb_k}), $\otimes$, simply means the direct product of two Hermitian matrices, $\mathbb{X}_{l_{1}\xi_{1},(s,k)}^{\rm (a)}$ and $\mathbb{Y}_{l_{2}\xi_{2}}^{\rm (s/b)}(\bm{k})$.

The momentum-space representation of the combined SAMB $\mathbb{Z}_{j}(\bm{k})$ is complete and orthonormal:
\begin{align}
  &
  \sum_{j} \left[\mathbb{Z}_{j}(\bm{k})\right]_{a_{1}a_{2}} \left[\mathbb{Z}_{j}(\bm{k}')\right]_{a_{3}a_{4}} = \delta_{a_{1},a_{4}} \delta_{a_{2},a_{3}} \delta_{\bm{k}, \bm{k}'},
  \\ &
  \frac{1}{N_{k}} \sum_{\bm{k}} \mathrm{Tr}\left[\mathbb{Z}_{i}(\bm{k}) \mathbb{Z}_{j}(\bm{k})\right] = \delta_{i,j}.
\end{align}

The real- and momentum-space combined SAMBs are transformed by the space group operation $U(g)$ given by Eq.~(\ref{eq_sym_oper_AO_R}) as
\begin{align}
&
U(g) \mathbb{Z}_{j}^{} U^{-1}(g) = \sum_{\gamma_{i}} \mathbb{Z}_{i}^{} D_{\gamma_{i} \gamma_{j}}^{\Gamma}(p),
\\ &
U(g) \mathbb{Z}_{j}^{}(\bm{k}) U^{-1}(g) = \sum_{\gamma_{i}} \mathbb{Z}_{i}^{}(\bm{k}) D_{\gamma_{i} \gamma_{j}}^{\Gamma}(p).
\end{align}

Following Eq.~(\ref{eq_c_samb_R}), the combined SAMBs in the graphene belonging to the identity $A_{1g}$ and symmetry breaking $A_{2u}$ irreps. up to 2nd-neighbor bond-cluster are summarized in Table~\ref{tbl_graphene_combined_samb}.
The schematic of each combined SAMB and their interpretation are given in Sec.~\ref{sec_graphene_E_field}.

\begin{figure*}[t]
  \begin{center}
  \includegraphics[width=0.98 \hsize]{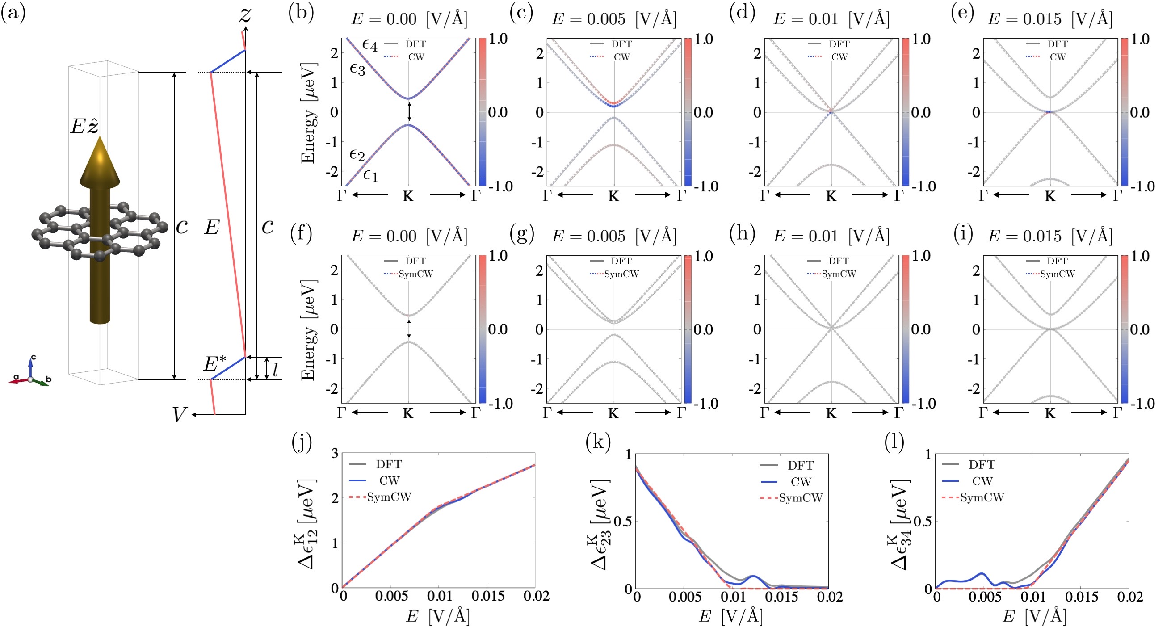}
  \vspace{5mm}
  \caption{
  \label{fig_graphene_E_dep}
 (a) Electric field along $c$ axis modeled with a zigzag potential, where $c = 20\, \AA$, $l = 1\, \AA$, and $E^{*} = -E (c-l)/l$.
  Electric field dependence of band structure of graphene near K point obtained by (b)-(e) the CW model and (f)-(i) the SymCW model represented by color dashed lines, where the gray solid lines represent the DFT band dispersion, and (b),(f) $E = 0.00$ [V/\AA], (c),(g) $E = 0.005$ [V/\AA], (d),(h) $E = 0.01$ [V/\AA], and (e),(i) $E = 0.015$ [V/\AA].
  The color map represents the $z$ component of spin.
  The time-reversal symmetry is broken in the DFT calculation and the CW model due to numerical error when $E > 0$, while it is strictly preserved in the SymCW model.
  The Fermi energy is set as the origin.
  Electric field dependence of the energy differences, (j) $\Delta\epsilon_{12}^{\rm K} = \epsilon_{2} - \epsilon_{1}$, (k) $\Delta\epsilon_{23}^{\rm K} = \epsilon_{3} - \epsilon_{2}$, and (l) $\Delta\epsilon_{34}^{\rm K} = \epsilon_{4} - \epsilon_{3}$ at K point.
  }
  \end{center}
\end{figure*}

\subsection{Symmetry-adapted closest Wannier model}
\label{sec_symcw_ham}

By utilizing the completeness and orthonormality of the combined SAMBs $\left\{\mathbb{Z}_{j}\right\}$, any operator $O$ can be expanded as a linear combination of $\mathbb{Z}_{j}$ as
\begin{align}
  O = \sum_{j}^{} o_{j} \mathbb{Z}_{j},
  \label{eq_O_samb}
\end{align}
where the expansion coefficient $o_{j}$ is determined by a simple matrix projection:
\begin{align}
  o_{j} &= \mathrm{Tr} \left[ \mathbb{Z}_{j} O \right] = \frac{1}{N_{k}} \sum_{\bm{k}} \mathrm{Tr} \left[ \mathbb{Z}_{j}(\bm{k}) O(\bm{k}) \right].
  \label{eq_oj}
\end{align}
The type of SAMB ($\mathbb{Z}$) considered in Eq.~(\ref{eq_O_samb}) is determined by the $\mathcal{P}$- and $\mathcal{T}$-parities of $O$.
For instance, when $O$ is $\mathcal{T}$-even polar (axial) quantity, $\mathbb{Z} = \mathbb{Q} (\mathbb{G})$, while $\mathbb{Z} = \mathbb{T} (\mathbb{M})$ when $O$ is $\mathcal{T}$-odd polar (axial) quantity.
Furthermore, if $O$ is characterized by $\Gamma$ irrep. and its component $\gamma$, only the coefficient of $\mathbb{Z}_{j}$ with $(\Gamma_{j}, \gamma_{j}) = (\Gamma, \gamma)$ becomes nonzero.

Since the CW Hamiltonian $H^{\rm CW}$ must be invariant for all the symmetry operations, we can express $H^{\rm CW}$ as a linear combination of $\mathbb{Z}_{j}$ belonging to the identity irrep. ($A$) as
\begin{align}
  &H^{\rm SymCW} = \sum_{j}^{\Gamma_{j} \in A} z_{j} \mathbb{Z}_{j},
  \label{eq_Hcw_samb}
  \\
  &z_{j} = \mathrm{Tr}\left[ \mathbb{Z}_{j} H^{\rm CW} \right].
  \label{eq_Hcw_samb_coeff}
\end{align}
It should be noted that $z_{j}$ is determined by a simple matrix projection with no iterative procedure by Eq.~(\ref{eq_Hcw_samb_coeff}) different from the previous method~\cite{PhysRevB.107.195118, RO_PRL_2022}, where $z_{j}$ is iteratively optimized so as to reproduce the DFT band dispersion, significantly reducing computational costs.
Since each SAMB corresponds to the CEF, SOC, or electron hoppings, as summarized in Table~\ref{tab_samb_hamiltonian}, we can numerically evaluate these model parameters by Eq.~(\ref{eq_Hcw_samb_coeff}).
Thus, our method is useful for identifying the primary contributions characterizing the material properties.
Furthermore, by considering a one-to-one correspondence between SAMBs and various external fields and responses~\cite{PhysRevB.104.054412, SH_JPSJ_review_2024} as summarized in Table~\ref{tab_samb_external_field_response}, our method also gives us insight into various unusual responses driven by hidden electronic multipole degrees of freedom~\cite{PhysRevB.104.054412, RO_PRL_2022}.

\section{Application to Graphene under Perpendicular Electric Field}
\label{sec_graphene_E_field}

\begin{figure}[t]
  \begin{center}
  \includegraphics[width=0.98 \hsize]{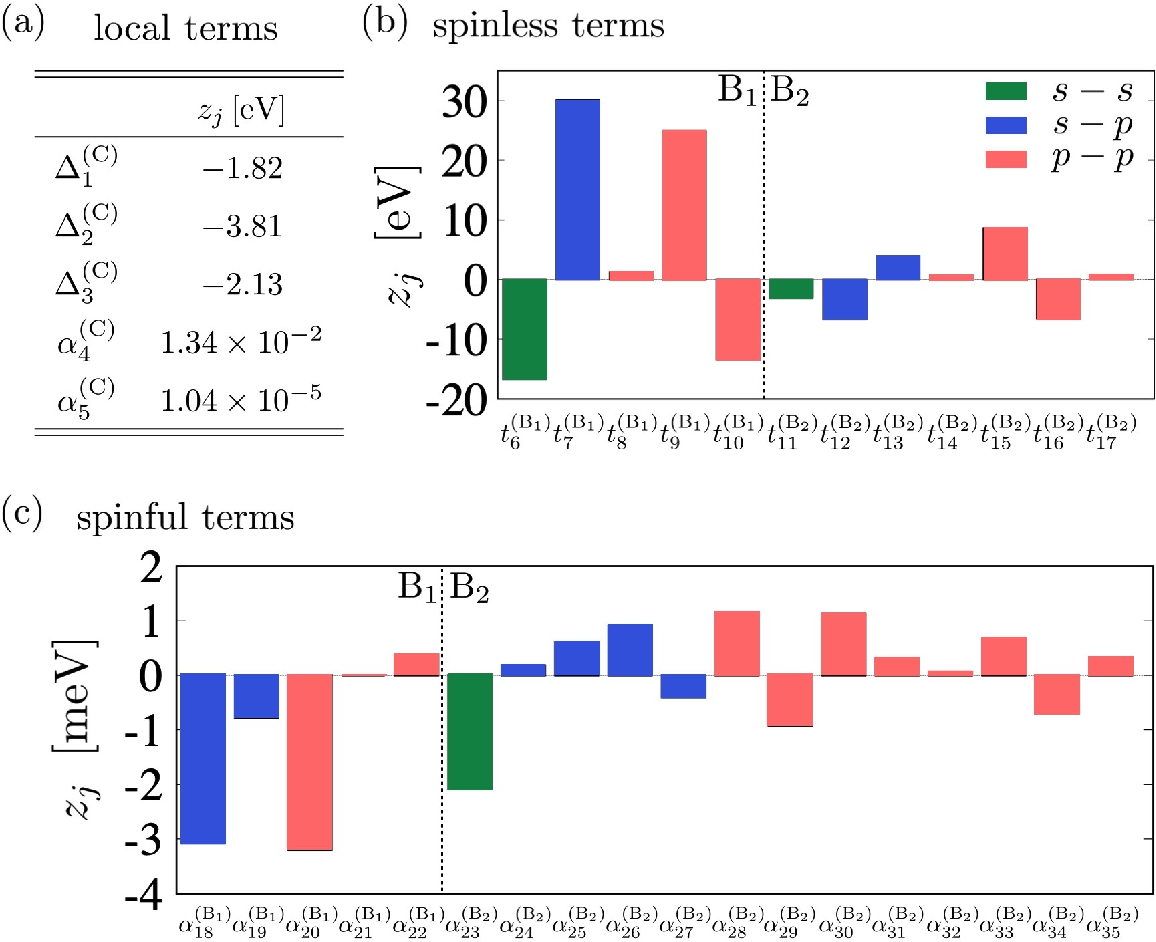}
  \vspace{5mm}
  \caption{
  \label{fig_graphene_z}
  Coefficients of each SAMB corresponding to (a) on-site terms, (b) spinless hopping terms, and (c) spinful hopping terms. 
  The green, blue, and red bars represent hoppings in the $\braket{s|s}$, $\braket{s|p}$, and $\braket{p|p}$ Hilbert spaces, respectively.
  }
  \end{center}
\end{figure}

\begin{figure*}[t]
  \begin{center}
  \includegraphics[width=0.98 \hsize]{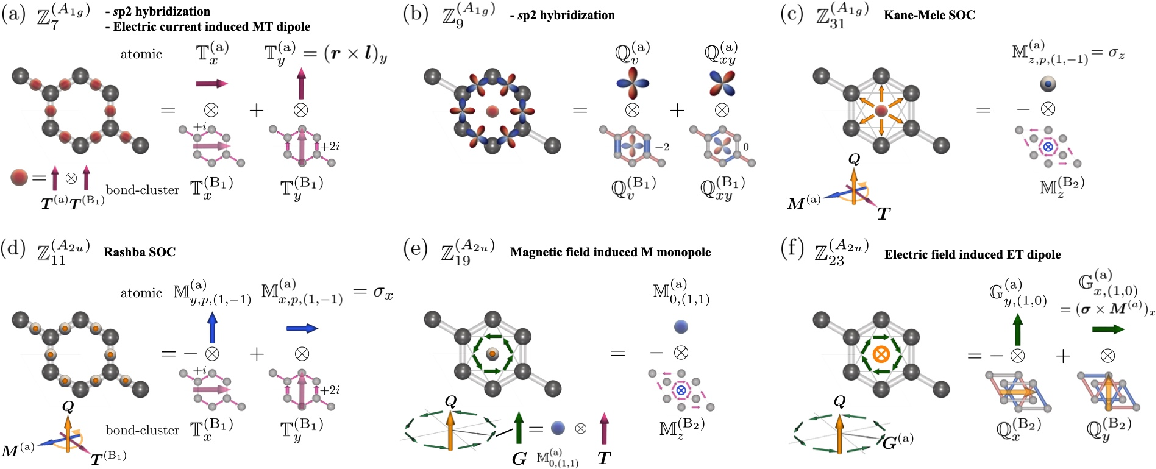}
  \vspace{5mm}
  \caption{
  \label{fig_graphene_combined_samb}
  Schematic picture of the combined SAMBs for graphene.
  The upper (a)-(c) panel and lower (d)-(f) panel represent the fully symmetric ($A_{1g}$ irrep.) terms and the symmetry breaking ($A_{2u}$ irrep.) terms in the D$_{\rm 6h}$ point group.
  (a) $\mathbb{Z}_{7}^{(A_{1g})}$ is the spinless nearest-neighbor real hoppings between $s$ and $(p_{x}, p_{y})$ orbitals and (b) $\mathbb{Z}_{9}^{(A_{1g})}$ is the spinless nearest-neighbor real hoppings among $(p_{x}, p_{y})$ orbitals.
  These two terms play an important role in constructing the $sp2$ hybridization.
  (c) $\mathbb{Z}_{31}^{(A_{1g})}$ corresponds to the 2nd-neighbor Kane-Mele SOC for $p$ orbitals, while (d) $\mathbb{Z}_{11}^{(A_{2u})}$ corresponds to the Bychkov-Rashba SOC for $p$ orbitals.
  (e) $\mathbb{Z}_{19}^{(A_{2u})}$ is the spinful 2nd-neighbor hoppings carrying the M monopole $\mathbb{M}_{0, (1,1)}^{\rm (a)}$, and (f) $\mathbb{Z}_{23}^{(A_{2u})}$ is the bond-orbital-spin interacting hoppings carrying the ET dipole $\mathbb{G}_{\mu, (1,0)}^{\rm (a)}$ ($\mu = x,y$).
  }
\end{center}
\end{figure*}

In this section, we demonstrate the SymCW modeling for monolayer graphene under an applied electric field perpendicular to the plane as an example.
For the DFT calculations, we have used the Quantum ESPRESSO~\cite{QE_Giannozzi_2009}.
We use the PBE exchange-correlation functional~\cite{Perdew_Burke_Ernzerhof_PBE_1996} and the full-relativistic optimized norm-conserving Vanderbilt (ONCV) pseudopotential~\cite{Hamann_PRB_2013}. 
The kinetic energy cutoff of the KS orbitals and the $\bm{k}$ grid, obtained by convergence tests, are set to 220 Ry and $(N_{1}, N_{2}, N_{3}) = (30, 30, 1)$, and the convergence threshold is set as $1\times10^{-13}$ Ry.

By using Quantum ESPRESSO~\cite{QE_Giannozzi_2009}, the electric field is considered in DFT calculations by saw-tooth potential as depicted in Fig.~\ref{fig_graphene_E_dep}(a)
\begin{align}
  V_{\rm ext}(z) = e E(z) z,
\end{align}
where $E(z) = E$ ($c > z > l$), $E(z) = E^{*} = -E (c-l)/l$ ($l \geq z \geq 0$), 
and $c = 20\, \AA$, and $l = 1\, \AA$.
Note that $V_{\rm ext}(z)$ satisfies the periodic boundary conditions.

The electric field $E$ dependence of band dispersions near K point obtained by the DFT calculation and the CW model are shown in Figs.~\ref{fig_graphene_E_dep}(b)-(e), where the color map represents the $z$ component of spin distribution in the momentum space.
As shown in Fig.~\ref{fig_graphene_E_dep}(b), when $E = 0$ the intrinsic SOC opens the gap at K point, where the obtained small gap $\Delta \epsilon_{23}^{\rm K} = \epsilon_{3} - \epsilon_{2} \sim 0.9$ $\mu$eV is consistent with the previous estimates 1 $\mu$eV~\cite{Min_PRB_2006, Yao_PRB_2007}, whereas it is much smaller than that obtained by the full-potential-based linearized augmented plane wave (LAPW) method considering $d$ orbitals~\cite{Gmitra_PRB_2009}.
The bands $\epsilon_{1}$ and $\epsilon_{2}$, and $\epsilon_{3}$ and $\epsilon_{4}$ are degenerate because of the spatial-
inversion and the time-reversal symmetries.

The electric field $E \neq 0$ breaks the spatial-inversion symmetry, lowering a point group symmetry from D$_{\rm 6h}$ to C$_{\rm 6v}$, and giving rise to the Bychkov-Rashba (BR) spin-splitting~\cite{Bychkov_Rashba_1984, KaneMele_PRL_2005, Rashba_PRB_2009}, $\Delta \epsilon_{12}^{\rm K} = \epsilon_{2}({\rm K}) - \epsilon_{1}({\rm K}) \sim 1$ $\mu$eV, when $E = 0.005$ [V/\AA] as shown in Fig.~\ref{fig_graphene_E_dep}(c).
Here, it should be noted that a Zeeman-like tiny gap $\Delta \epsilon_{34}^{\rm K} = \epsilon_{4}({\rm K}) - \epsilon_{3}({\rm K}) \sim 0.1$ $\mu$eV appears near K point with nonzero $z$ component of spin due to numerical error.
Consequently, the broken time-reversal symmetry occurs unfavorably in both the DFT calculation and CW model.
As the electric field is increased, the band topologies near K point drastically change as shown in Figs.~\ref{fig_graphene_E_dep}(c), (d), and (e)~\cite{Gmitra_PRB_2009}.
The electric field dependence of the gaps at K point is shown in Figs.~\ref{fig_graphene_E_dep}(j)-(l).
The gaps, $\Delta\epsilon_{23}^{\rm K}$ above $E = 0.01$ and $\Delta\epsilon_{34}^{\rm K}$ below $E = 0.01$, are slightly opened in the band dispersions of the DFT calculation and the CW model.

\subsection{Symmetrization of the CW model}

Here, we demonstrate that the unexpected small gap at K point is closed by using the SymCW modeling, thereby restoring the broken symmetries.
Since the electric field belongs to the $A_{2u}$ irrep. in D$_{\rm 6h}$ point group, by considering the combined SAMBs with identity $A_{1g}$ and the symmetry breaking $A_{2u}$ irreps., the SymCW Hamiltonian is obtained by
\begin{align}
  H^{\rm SymCW} &= H^{(A_{1g})} + H^{(A_{2u})},
  \label{eq_H_symcw}
  \\
  H^{(A_{1g})} &= H_{\rm CEF}^{\rm (C)} + H_{\rm SOC}^{\rm (C)} + \sum_{n = 1}^{N_{\rm B}} \left(H_{t}^{\rm (B_{n})} + H_{t}^{\rm (B_{n})'}\right),
  \\
  H_{\rm CEF}^{\rm (C)} &= \Delta_{1}^{\rm (C)} \mathbb{Z}_{1}^{\rm (A_{1g})} + \Delta_{2}^{\rm (C)} \mathbb{Z}_{2}^{\rm (A_{1g})} + \Delta_{3}^{\rm (C)} \mathbb{Z}_{3}^{\rm (A_{1g})},
  \\
  H_{\rm SOC}^{\rm (C)} &= \alpha_{4}^{\rm (C)} \mathbb{Z}_{4}^{\rm (A_{1g})} + \alpha_{5}^{\rm (C)} \mathbb{Z}_{5}^{\rm (A_{1g})},
  \\
  H_{t}^{\rm (B_{1})} &= \sum_{j=6}^{10} t_{j}^{\rm (B_{1})} \mathbb{Z}_{j}^{\rm (A_{1g})}, \quad H_{t}^{\rm (B_{2})} = \sum_{j=11}^{17} t_{j}^{\rm (B_{2})} \mathbb{Z}_{j}^{\rm (A_{1g})},
  \\
  H_{t}^{\rm (B_{1})'} &= \sum_{j=18}^{22} \alpha_{j}^{\rm (B_{1})} \mathbb{Z}_{j}^{\rm (A_{1g})}, \quad H_{t}^{\rm (B_{2})'} = \sum_{j=23}^{35} \alpha_{j}^{\rm (B_{2})} \mathbb{Z}_{j}^{\rm (A_{1g})},
\end{align}
where $H^{(A_{1g})}$ and $H^{(A_{2u})}$ represent the $A_{1g}$ and $A_{2u}$ terms, respectively.
$H^{(A_{1g})}$ is decomposed into the on-site CEF and SOC terms within the unit cell, and the spinless $H_{t}^{\rm (B_{n})}$ and spinful $H_{t}^{\rm (B_{n})'}$ hopping terms.
By taking up to the $N_{\rm B} = 108$ bond-cluster with the bond length $|\bm{b}| = 25.60\, \AA$, we consider 8722 combined SAMBs in total.
When $E = 0$, the mean absolute error between the energy eigenvalues of the CW model and the SymCW model is 
\begin{align}
\frac{1}{N_{k} N_{\rm W}} \sum_{n\bm{k}} |\epsilon_{n\bm{k}}^{\rm CW} - \epsilon_{n\bm{k}}^{\rm SymCW}| \simeq 0.013\, {\rm meV}.
\end{align}
The comparison of the energy dispersions is shown in Figs.~\ref{fig_graphene_cw_band}(b), (c), and (d), where the orbital characters of the obtained electronic states well reproduce those of the CW model.
Moreover, the slight lifting of the band degeneracy $\Delta \epsilon_{34}^{\rm K} \neq 0$ and the broken time-reversal symmetry in the DFT calculation and the CW model is fully recovered as shown in Figs.~\ref{fig_graphene_E_dep}(f)-(i).
In addition, the unexpected slight opening of gaps, $\Delta\epsilon_{23}^{\rm K}$ above $E = 0.01$ and $\Delta\epsilon_{34}^{\rm K}$ below $E = 0.01$, are closed in the SymCW model as shown in Figs.~\ref{fig_graphene_E_dep}(j)-(l).
This is one of the most significant results of this paper.
In this approach, the CW Hamiltonian is reconstructed by using only the SAMBs with identity irrep., which restores the broken symmetries in the DFT calculations and CW models.
Similar results for bcc Nb and the spin-orbit coupled metal monolayer MoS$_{2}$ are provided in the Supplemental Materials.

\subsection{Physical interpretation of SAMBs ($E = 0$)}

Next, we present the physical interpretation of each SAMB.
When $E = 0$, within the 2nd-neighbor bond-cluster, there are 35 SAMBs belonging to $A_{1g}$, in which 3 CEF ($\Delta_{1}^{\rm (C)} \sim \Delta_{3}^{\rm (C)}$) and 2 SOC ($\alpha_{4}^{\rm (C)}, \alpha_{5}^{\rm (C)}$) parameters, and 12 spinless ($t_{6}^{\rm (B_{1})} \sim t_{10}^{\rm (B_{1})} , t_{11}^{\rm (B_{2})} \sim t_{17}^{\rm (B_{2})}$) and 18 spinful ($\alpha_{18}^{\rm (B_{1})} \sim \alpha_{22}^{\rm (B_{1})}, \alpha_{23}^{\rm (B_{2})} \sim \alpha_{35}^{\rm (B_{2})}$) hopping parameters.
They are evaluated by Eq.~(\ref{eq_Hcw_samb_coeff}) as summarized in Fig.~\ref{fig_graphene_z}, and the corresponding combined SAMBs are summarized in Table~\ref{tbl_graphene_combined_samb}.

The CEF terms are given by
\begin{align}
  &\mathbb{Z}_{1}^{(A_{1g})} = \mathbb{Q}_{0,s}^{\rm (a)}\otimes\mathbb{Q}_{0}^{\rm (C)},
  \label{eq_z1_A1g} \\
  &\mathbb{Z}_{2}^{(A_{1g})} = \mathbb{Q}_{0,p}^{\rm (a)}\otimes\mathbb{Q}_{0}^{\rm (C)},
  \label{eq_z2_A1g} \\
  &\mathbb{Z}_{3}^{(A_{1g})} = \mathbb{Q}_{u}^{\rm (a)}\otimes\mathbb{Q}_{0}^{\rm (C)}.
  \label{eq_z3_A1g}
\end{align}
By using their matrix expressions given in Table~\ref{tab_a_samb}, the CEF Hamiltonian is obtained as
\begin{align}
&H_{\rm CEF} =
\begin{pmatrix}
  \Delta_{s} & 0 & 0 & 0 \\
0 & \Delta_{p_{x}} & 0 & 0 \\
0 & 0 & \Delta_{p_{y}} & 0 \\
0 & 0 & 0 & \Delta_{p_{z}} \\
\end{pmatrix}
\otimes \sigma_{0} \otimes \rho_{0},
\\
& \Delta_{s} = \frac{1}{2} \Delta_{1}^{\rm (C)} = -0.912\, {\rm eV}, \\
& \Delta_{p_{x}} = \Delta_{p_{y}} = \frac{\sqrt{2}\Delta_{2}^{\rm (C)} - \Delta_{3}^{\rm (C)}}{2\sqrt{6}} = -0.664\, {\rm eV}, \\
& \Delta_{p_{z}} = \frac{\sqrt{2}\Delta_{2}^{\rm (C)} + 2 \Delta_{3}^{\rm (C)}}{2\sqrt{6}} = -1.97\, {\rm eV},
\end{align}
where $\rho_{0}$ is a 2$\times$2 identity matrix in the sublattice space, (C$_{1}$, C$_{2}$).
The difference between $s$ and $(p_{x}, p_{y})$ orbitals, $\Delta_{s-p_{x}} = \Delta_{s} - \Delta_{p_{x}} = -0.248$ eV is much smaller than that between $s$ and $p_{z}$ orbitals, $\Delta_{s-p_{z}} = \Delta_{s} - \Delta_{p_{z}} = 1.06$ eV.
Because of the hexagonal structure, there is a difference between $(p_{x},p_{y})$ and $p_{z}$ orbitals as $\Delta_{p_{x}-p_{z}} = \Delta_{p_{x}} - \Delta_{p_{z}} = 1.31$ eV.

The atomic SOC terms are given by
\begin{align}
&\mathbb{Z}_{4}^{(A_{1g})} = \mathbb{Q}_{0,(1,1)}^{\rm (a)}\otimes\mathbb{Q}_{0}^{\rm (C)},
\\
&\mathbb{Z}_{5}^{(A_{1g})} = \mathbb{Q}_{u,(1,-1)}^{\rm (a)}\otimes\mathbb{Q}_{0}^{\rm (C)}.
\end{align}
By using their matrix expressions given in Table~\ref{tab_a_samb}, the atomic SOC Hamiltonian is obtained as
\begin{align}
& H_{\rm SOC} = \frac{\xi_{0}}{2} \bm{l}\cdot\bm{\sigma} \otimes \rho_{0} +\frac{\xi_{u}}{2} (2l_{z} \sigma_{z} - l_{x} \sigma_{x} - l_{y} \sigma_{y}) \otimes \rho_{0},
\\
& \xi_{0} = \frac{1}{\sqrt{6}} \alpha_{4}^{\rm (C)} = 5.48\, {\rm meV},
\\
& \xi_{u} = \frac{1}{2\sqrt{3}} \alpha_{5}^{\rm (C)} = 3.01 \times10^{-3}\, {\rm meV},
\end{align}
where $\bm{l}$ and $\bm{\sigma}/2$ represent the dimensionless orbital and spin angular-momentum operators, respectively.
$\xi_{0}$ corresponds to the atomic SOC constant, and its amplitude $\xi_{0} = 5.48$ meV is reasonably consistent with the value 6 meV obtained by fitting the TB model to the DFT calculation~\cite{Min_PRB_2006}.
It is important to note that the above model parameters can be obtained through non-iterative calculations in our method, whereas the conventional fitting approach requires iterative calculations to determine the parameters.
The term $\xi_{u}$ is induced by the hexagonal CEF potential, and its amplitude is negligibly smaller than that of the spherical potential term $\xi_{0}$.

\begin{figure*}[t]
  \begin{center}
  \includegraphics[width=0.98 \hsize]{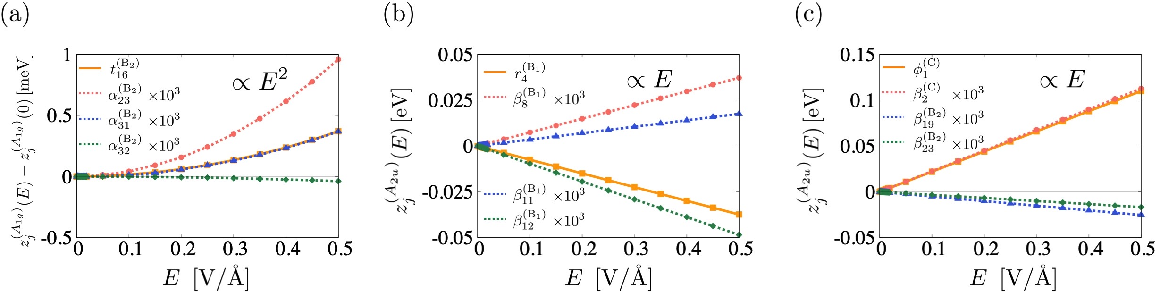}
  \vspace{5mm}
  \caption{
  \label{fig_graphene_E_dep_samb}
  Electric field dependence of the coefficients for each SAMBs.
  (a) $A_{1g}$ irrep. terms measured from their values when $E = 0$, and (b), (c) $A_{2u}$ irrep. terms.
  }
  \end{center}
\end{figure*}

Among the spinless hoppings, the most significant contributions are given by
\begin{align}
&\mathbb{Z}_{7}^{(A_{1g})} = \frac{1}{\sqrt{2}} \left( \mathbb{T}_{x}^{\rm (a)}\otimes\mathbb{T}_{x}^{\rm (B_{1})} + \mathbb{T}_{y}^{\rm (a)}\otimes\mathbb{T}_{y}^{\rm (B_{1})} \right),
\label{eq_z7_A1g} \\
&\mathbb{Z}_{9}^{(A_{1g})} = \frac{1}{\sqrt{2}} \left( \mathbb{Q}_{v}^{\rm (a)}\otimes\mathbb{Q}_{v}^{\rm (B_{1})} + \mathbb{Q}_{xy}^{\rm (a)}\otimes\mathbb{Q}_{xy}^{\rm (B_{1})} \right).
\label{eq_z9_A1g}
\end{align}
The schematics of them are shown in Figs.~\ref{fig_graphene_combined_samb}(a) and (b), respectively.
The weight of $\mathbb{Z}_{7}^{(A_{1g})}$ is $t_{7}^{\rm (B_{1})} = 30.1$ eV which is the most dominant contribution, and that of $\mathbb{Z}_{9}^{(A_{1g})}$ is $t_{9}^{\rm (B_{1})} = 24.9$ eV.
We have confirmed that the magnitude of the hopping parameters decreases for further neighbor hoppings.
Here, $\mathbb{T}_{\mu}^{\rm (a)} \propto \bm{t} = (\bm{r} \times \bm{l})_{\mu}$ ($\mu = x,y$) in $\mathbb{Z}_{7}^{(A_{1g})}$ is the atomic MT dipole ($\bm{r}$ is the atomic position operator) defined in the $\braket{s|p}$ Hilbert space as shown in Table~\ref{tab_a_samb}, and $\mathbb{T}_{\mu}^{\rm (B_{1})}$ ($\mu = x,y$) is the nearest-neighbor bond-cluster MT dipole as shown in Table~\ref{tbl_graphene_sb_samb}.
On the other hand, $\mathbb{Q}_{v}^{\rm (a)} \propto x^{2}-y^{2}$ and $\mathbb{Q}_{xy}^{\rm (a)} \propto xy$ in $\mathbb{Z}_{9}^{(A_{1g})}$ are the atomic E quadrupole defined in the $\braket{p|p}$ Hilbert space, and $\mathbb{Q}_{v}^{\rm (B_{1})}$ and $\mathbb{Q}_{xy}^{\rm (B_{1})}$ are the nearest-neighbor bond-cluster E quadrupoles.
Therefore, $\mathbb{Z}_{7}^{(A_{1g})}$ and $\mathbb{Z}_{9}^{(A_{1g})}$ correspond to the spinless nearest-neighbor real hoppings between $s$ and $(p_{x}, p_{y})$ orbitals and among $(p_{x}, p_{y})$ orbitals, respectively, giving rise to the deep-lying $\sigma$ bands by $sp2$ hybridization.
Much large values of $t_{7}^{\rm (B_{1})} = 30.1$ eV and $t_{9}^{\rm (B_{1})} = 24.9$ eV are consistent with the robustness of the two-dimensional structure of graphene.

It should be noted that there is also the spinful term $\mathbb{Z}_{18}^{(A_{1g})}$ that has the same form as Eq.~(\ref{eq_z7_A1g}), where the spinless MT dipole $\mathbb{T}_{\mu}^{\rm (a)}$ is replaced by the spinful MT dipole $\mathbb{T}_{\mu, (1,0)}^{\rm (a)} \propto (\bm{r} \times \bm{\sigma})_{\mu}$ ($\mu = x,y$).
As shown in Fig.~\ref{fig_graphene_z}(c), the weight of $\mathbb{Z}_{18}^{(A_{1g})}$, $\alpha_{18} = -3.08$ meV, is also large among the spinful terms.
Then, $\mathbb{Z}_{18}^{(A_{1g})}$ corresponds to the spinful nearest-neighbor imaginary hoppings between $s$ and $(p_{x}, p_{y})$ orbitals.

On the other hand, in the spinful 2nd-neighbor hoppings, there is a term corresponding to the Kane-Mele SOC given by~\cite{KaneMele_PRL_2005}
\begin{align}
&\mathbb{Z}_{31}^{(A_{1g})} = -\mathbb{M}_{z,p,(1,-1)}^{\rm (a)}\otimes\mathbb{M}_{z}^{\rm (B_{2})},
\label{eq_z31_A1g}
\end{align}
where $\mathbb{M}_{z,p,(1,-1)}^{\rm (a)} = \sigma_{z}$ is the $z$ component of spin defined in the diagonal part of the $\braket{p|p}$ Hilbert space, and $\mathbb{M}_{z}^{\rm (B_{2})}$ is nothing but the Haldane's magnetic flux corresponding to the vortexlike imaginary hopping in 2nd-neighbor C$_{1}$-C$_{1}$ or C$_{2}$-C$_{2}$ bonds as shown in Fig.~\ref{fig_graphene_combined_samb}(c)~\cite{Haldane_PRL_1988}.
The weight of $\mathbb{Z}_{31}^{(A_{1g})}$ is $\alpha_{31}^{\rm (B_{2})} = 0.306$ meV.
The momentum space representation of Eq.~(\ref{eq_z31_A1g}) at $\pm$K point is given by
\begin{align}
  &\mathbb{Z}_{31}^{(A_{1g})}(\pm {\rm K}) = \pm \lambda_{\rm KM}  \sigma_{z} \otimes \rho_{z}, \\
  &\lambda_{\rm KM} = -\frac{\sqrt{6}}{4} \alpha_{31}^{\rm (B_{2})} = 0.187\, {\rm meV},
\end{align}
where $\rho_{z}$ is the Pauli matrix for the sublattice space.
Thus, $\mathbb{M}_{z}^{\rm (B{2})}$ acts as an effective magnetic field during the hopping process at the bond center, with its sign changed depending on the hopping pass, C$_{1}$-C$_{1}$ or C$_{2}$-C$_{2}$.
Thus, $\mathbb{Z}_{31}^{(A_{1g})}$ is a primarily contribution to the microscopic origin of the band splitting at K point as shown in Fig.~\ref{fig_graphene_E_dep}(b).
Note that the band gap size is not determined by this single parameter, however, there are multiple contributions from further neighbor hoppings having the same form as $\mathbb{Z}_{31}^{(A_{1g})}$.
Indeed, there are three such combined SAMBs, $\mathbb{Z}_{16}^{(A_{1g})}$, $\mathbb{Z}_{23}^{(A_{1g})}$, and $\mathbb{Z}_{32}^{(A_{1g})}$, where $\mathbb{M}_{z,p,(1,-1)}^{\rm (a)}$ in Eq.~(\ref{eq_z31_A1g}) is replaced by the orbital M dipole $\mathbb{M}_{z}^{\rm (a)}$, the spin M dipole $\mathbb{M}_{z,s,(1,-1)}^{\rm (a)}$ defined in $\braket{s|s}$, and the anisotropic M dipole $\mathbb{M}_{z,a,(1,1)}^{\rm (a)}$ defined in $\braket{p|p}$.
As shown in Figs.~\ref{fig_graphene_z}(b) and (c), the amplitude of the coefficient for the spinless $\mathbb{Z}_{16}^{(A_{1g})}$ is 1000 times larger than those of the spinful $\mathbb{Z}_{23}^{(A_{1g})}$, $\mathbb{Z}_{31}^{(A_{1g})}$, and $\mathbb{Z}_{32}^{(A_{1g})}$, because the relativistic effect is small in graphene.

\subsection{Interpretation of SAMBs ($E \neq 0$)}

When $E \neq 0$ the inversion symmetry is broken, then the symmetry breaking term $H^{(A_{2u})}$ appears in Eq.~(\ref{eq_H_symcw}).
Within the 2nd-neighbor bond-cluster, there are 28 SAMBs belonging to $A_{2u}$, in which 1 CEF ($\phi_{1}^{(\rm C)}$) and 1 SOC ($\beta_{2}^{(\rm C)}$) parameters, and 5 spinless ($r_{3}^{\rm (B_{1})} \sim r_{4}^{\rm (B_{1})}, r_{5}^{\rm (B_{2})} \sim r_{7}^{\rm (B_{2})}$) and 21 spinful ($\beta_{8}^{\rm (B_{1})} \sim \beta_{15}^{\rm (B_{1})}, \beta_{16}^{\rm (B_{2})} \sim \beta_{28}^{\rm (B_{2})}$) hopping parameters.
The combined SAMBs are summarized in the lower panel of Table~\ref{tbl_graphene_combined_samb}.
In particular, since the spinless and spinful atomic E dipoles, $\mathbb{Q}_{z}^{\rm (a)} \propto z$ and $\mathbb{Q}_{z, (1,0)}^{\rm (a)} \propto (\bm{\sigma} \times \bm{t})_{z}$, have the same symmetry as the electric field, they appear as the on-site 
symmetry breaking term as
\begin{align}
&\mathbb{Z}_{1}^{(A_{2u})} = \mathbb{Q}_{z}^{\rm (a)}\otimes\mathbb{Q}_{0}^{\rm (C)},
\label{eq_z1_A2u}
\\
&\mathbb{Z}_{2}^{(A_{2u})} = \mathbb{Q}_{z, (1,0)}^{\rm (a)}\otimes\mathbb{Q}_{0}^{\rm (C)}.
\label{eq_z2_A2u}
\end{align}

On the other hand, there is a spinful hopping term corresponding to the Bychkov-Rashba SOC for $p$ orbitals appeared by the perpendicular electric field:
\begin{align}
&\mathbb{Z}_{11}^{(A_{2u})} =  \frac{1}{\sqrt{2}}\left(\mathbb{M}_{x,p,(1,-1)}^{\rm (a)}\otimes\mathbb{T}_{y}^{\rm (B_{1})} - \mathbb{M}_{y,p,(1,-1)}^{\rm (a)}\otimes\mathbb{T}_{x}^{\rm (B_{1})}\right),
\label{eq_z11_A2u}
\end{align}
where $\mathbb{M}_{\mu,p,(1,-1)}^{\rm (a)} = \sigma_{\mu}$ is the $\mu=x,y$ component of spin defined in the diagonal part of the $\braket{p|p}$ Hilbert space, and $\mathbb{T}_{\mu}^{\rm (B_{1})}$ is the nearest-neighbor bond-cluster MT dipole.
The Fourier transform of Eq.~(\ref{eq_z11_A2u}) near $\pm$K point is given by
\begin{align}
  &\mathbb{Z}_{11}^{(A_{2u})}(\pm {\rm K}) = \lambda_{\rm BR} \left[(\mp \sigma_{y} \otimes \rho_{x} + \sigma_{x} \otimes \rho_{y}) \right.
  \cr & \left. \qquad\qquad\qquad + \sqrt{3} (\pm \sigma_{x} \otimes \rho_{x} + \sigma_{y} \otimes \rho_{y})\right], \\
  &\lambda_{\rm BR} = \frac{1}{8} 
  \beta_{11}^{(B_{1})},
\end{align}
where $\rho_{x}$ and $\rho_{y}$ are the Pauli matrices for the sublattice space.
Therefore, $\mathbb{Z}_{11}^{(A_{2u})}$ is a primarily contribution to the microscopic origin of the Rashba splitting near $\pm$K point as shown in Figs.~\ref{fig_graphene_E_dep}(b), (c), and (d).
Note that the band gap size is not determined by this single parameter, however, there are multiple contributions from further neighbor hoppings which have same form as $\mathbb{Z}_{11}^{(A_{2u})}$.
Indeed, there are three such combined SAMBs, $\mathbb{Z}_{4}^{(A_{2u})}$, $\mathbb{Z}_{8}^{(A_{2u})}$, and $\mathbb{Z}_{12}^{(A_{2u})}$, where $\mathbb{M}_{\mu,p,(1,-1)}^{\rm (a)}$ in Eq.~(\ref{eq_z11_A2u}) is replaced by the spinless M dipole $\mathbb{M}_{\mu}^{\rm (a)}$, the spin M dipole $\mathbb{M}_{\mu,s,(1,-1)}^{\rm (a)}$ defined in $\braket{s|s}$, and the anisotropic M dipole $\mathbb{M}_{\mu,a,(1,1)}^{\rm (a)}$ defined in $\braket{p|p}$.

Additionally, there are 2nd-neighbor spinful hoppings including the atomic M monopole $\mathbb{M}_{0,(1,1)}^{\rm (a)}$ and the ET dipole $\mathbb{G}_{\mu,(1,0)}^{\rm (a)}$ given by
\begin{align}
&\mathbb{Z}_{19}^{(A_{2u})} = \mathbb{M}_{0,(1,1)}^{\rm (a)}\otimes\mathbb{M}_{z}^{\rm (B_{2})},
\label{eq_z19_A2u}
\\
&\mathbb{Z}_{23}^{(A_{2u})} =  \frac{1}{\sqrt{2}}\left(\mathbb{G}_{x,(1,0)}^{\rm (a)}\otimes\mathbb{Q}_{y}^{\rm (B_{2})} - \mathbb{G}_{y,(1,0)}^{\rm (a)}\otimes\mathbb{Q}_{x}^{\rm (B_{2})}\right).
\label{eq_z23_A2u}
\end{align}
The schematics of them are shown in Figs.~\ref{fig_graphene_combined_samb}(e) and (f), respectively.
In this way, the symmetry-breaking terms are easily clarified in terms of the SAMB.

Let us discuss the electric field dependence of the coefficients for each SAMB.
Since the irreducible decomposition of $E^{2n}$ and $E^{2n+1}$ belong to $A_{1g}$ and $A_{2u}$, respectively, the CW Hamiltonian $H^{\rm CW}$ can be expanded with respective to the electric field $E$ as
\begin{align}
H^{\rm CW}(E) = \sum_{n=0} \left[H^{\rm CW}(A_{1g}, n) E^{2n} + H^{\rm CW}(A_{2u}, n) E^{2n+1}\right].
\end{align}
Here, $H^{\rm CW}(A_{1g}, n)$ and $H^{\rm CW}(A_{2u}, n)$ belong to the $A_{1g}$ and $A_{2u}$ irreps, respectively.
Note that Eq.~(\ref{eq_Hcw_samb_coeff}) becomes nonzero when the identity $A_{1g}$ irrep. is contained in the irreducible decomposition of the product of $\mathbb{Z}_{j}$ and $H^{\rm CW}(E)$.
Then, the coefficients $z_{j}^{(A_{1g})}(E)$ and $z_{j}^{(A_{2u})}(E)$ are obtained by
\begin{align}
  &z_{j}^{(A_{1g})}(E) = \sum_{n=0} \mathrm{Tr}\left[ \mathbb{Z}_{j}^{(A_{1g})} H^{\rm CW}(A_{1g}, n) \right] E^{2n},
  \\ &
  z_{j}^{(A_{2u})}(E) = \sum_{n=0} \mathrm{Tr}\left[ \mathbb{Z}_{j}^{(A_{2u})} H^{\rm CW}(A_{2u}, n) \right] E^{2n+1}.
\end{align}
Consequently, when the magnitude of electric field is small, $z_{j}^{(A_{1g})}(E) - z_{j}^{(A_{1g})}(0) \propto E^{2}$ and $z_{j}^{(A_{2u})}(E) \propto E$, which is consistent with the numerical results as shown in Figs.~\ref{fig_graphene_E_dep_samb}(a)-(c).

\subsection{Expected responses}
\label{sec_responses}

In order to obtain deeper insights into physical responses, SymCW modeling based on SAMB is also useful. 
Let us first begin with the case without perpendicular electric field.
For example, let us focus on Eq.~(\ref{eq_z7_A1g}).
Since the MT dipole has the same symmetry as the electric current $\bm{J}$ and the composite field of the electric and magnetic fields $(\bm{E}\times\bm{H})$ as shown in Table~\ref{tab_samb_external_field_response}, $J_{\mu}$ or $(\bm{E}\times\bm{H})_{\mu}$ induced atomic orbital and spinful MT moments $\braket{\mathbb{T}_{\mu}^{\rm (a)}}$ and $\braket{\mathbb{T}_{\mu, (1,0)}^{\rm (a)}}$ can be realized.
Recently, the photo-induced MT dipole moment has been discussed based on Floquet formalism~\cite{SH_MT_Flouquet_2024}.
As mentioned in~\cite{SH_MT_Flouquet_2024}, in a single atom limit, the photo-induced $\braket{\mathbb{T}_{\mu}^{\rm (a)}}$ and $\braket{\mathbb{T}_{\mu, (1,0)}^{\rm (a)}}$ show explicit different model parameter dependence.
The former and latter are related to the atomic energy difference between $s$ and $p_{\mu}$ $\Delta_{s-p_{\mu}}$ and the atomic SOC constant $\xi_{0}$, respectively.
From our results, $\Delta_{s-p_{\mu}} = -0.248$ eV is much larger than $\xi_{0} = 5.48$ meV, suggesting that a photo-induced pure atomic orbital MT dipole is more likely to be observed experimentally than a spinful MT dipole.

On the other hand, under perpendicular electric field, there appears the Bychkov-Rashba SOC given by Eq.~(\ref{eq_z11_A2u}).
Then, the Rashba Edelstein effect $\braket{M_{x}} = \alpha^{(J)}_{x;y} E_{y}$, where $\alpha^{(J)}_{x;y}$ is the linear response coefficient, can be realized~\cite{PhysRevB.89.075422, PhysRevLett.119.196801,ghiasiChargetoSpinConversionRashba2019,PhysRevB.104.235429}.
Moreover, from Eq.~(\ref{eq_z19_A2u}), we can easily predict that the atomic M monopole can be induced by applying the magnetic field perpendicular to the plane.
Similarly, from Eq.~(\ref{eq_z19_A2u}), the electric field induced ET dipoles is predicted to be realized, which is described by $\braket{G_{x}} = d_{x;y} E_{y}$ where $d_{x;y}$ is the linear response coefficient.

Furthermore, following the systematic analysis method proposed in Refs.~\cite{PhysRevB.102.144441, RO_JPSJ_2022} for the expectation value of an operator $O$ and the response functions of the output $A$ and input $B$ operators, we can extract essential parameters in a systematic manner by analyzing the indicators such as
\begin{align}
  \Gamma^{i} & = \mathrm{Tr} \left[O H^i\right], \\
  \Gamma_{\mu ; \alpha}^{i j} & = \mathrm{Tr}\left[A_\mu H^i B_\alpha H^j\right], \\
  \Gamma_{\mu ; \alpha, \beta}^{i j k} & = \mathrm{Tr}\left[A_\mu H^i B_\alpha H^j B_\beta H^k\right],
\end{align}
where $H^{i}$ is the $i$th power of the Hamiltonian matrices.
By using the SymCW Hamiltonian given by Eq.~(\ref{eq_H_symcw}), the trace in the above indicators is regarded as selecting the identity irrep. in the irreducible decomposition of the product of $O$ and $H^{i}$ or $A$, $B$, and $H^{i}$.
Thus, the combination of operators that yields the identity irrep. represents the essential parameters, and the expectation value or response functions becomes finite when the corresponding indicator is nonzero.

Given these results, SymCW modeling provides quantitative guidelines for future material design that go beyond the conventional phenomenological approach based on the group theory and the standard DFT-based Wannier TB modeling.
In this way, SymCW modeling enables the prediction of unusual orderings and physical responses, and provides us with deeper insights into their underlying microscopic mechanisms.

\section{Summary}
\label{sec_Summary}

In this paper, we have developed a symmetry-adapted closest Wannier (SymCW) modeling procedure.
Since the symmetry properties of the closest Wannier functions are common to those of the original atomic orbitals, we introduced symmetry-adapted multipole bases (SAMBs) as the complete orthonormal basis set in the Hilbert space of the closest Wannier functions.
Consequently, any physical operator can be expressed as a linear combination of SAMBs.
In particular, the SymCW model is obtained by reconstructing the closest Wannier Hamiltonian as a linear combination of SAMBs belonging to the identity irreducible representation. 
Therefore, it is useful to remove unfavorable symmetry breakings caused by numerical errors in the DFT calculation and the closest Wannier model.
The linear coefficients of each SAMB correspond to the model parameters related to the crystalline electric fields, spin-orbit coupling, and electron hoppings.
Unlike the previous methods~\cite{PhysRevB.107.195118, RO_PRL_2022}, these coefficients are determined through a simple matrix projection without any iterative process, significantly reducing the computational costs.
Thus, the SymCW modeling allows accurate, non-iterative determination of model parameters, which are difficult to extract in standard DFT-based tight-binding approaches.
Moreover, our method provides deeper insights into various unconventional responses driven by electronic multipole degrees of freedom hidden in the closest Wannier Hamiltonian. 

We have demonstrated key applications of our method by modeling monolayer graphene under a perpendicular electric field as a simplest example.
We have shown that the unfavorable time-reversal symmetry breaking appears under electric field in both the DFT calculation and closest Wannier model.
It is remedied by the SymCW model, ensuring that the SymCW model rigorously preserves the symmetry of the system.
This is one of the major advantages of our method.
Additionally, we have explained physical interpretation for each SAMB appearing in the SymCW model and emphasized the usefulness of our method to predict unusual responses quantitatively in terms of SAMBs.

Although we have applied our method to one-body tight-binding modeling, it can also be extended to reconstruct two-body interactions, such as density-density one, magnetic exchange couplings including Dzyaloshinskki-Moriya type~\cite{PhysRevB.101.224419, PhysRevB.104.134420,PhysRevB.105.014404}, as a linear combination of products of two SAMBs.
Furthermore, our method can also be used to symmetrize and reconstruct the phonon dynamical matrix as a linear combination of SAMBs~\cite{doi:10.7566/JPSJ.92.012001,doi:10.7566/JPSJ.92.023601}.
These are topics for future work.
In this way, the present method facilitates efficient and high-precision modeling, with the potential to guide future material designs beyond conventional group-theory-based phenomenological approach and standard DFT-based Wannier tight-binding modeling.

The SAMBs can be automatically generated by using the open-source Python library MultiPie (https://github.com/CMT-MU/MultiPie), and the SymCW modeling method is implemented in the open-source Python library SymClosestWannier (https://github.com/CMT-MU/SymClosestWannier) to determine the model parameters from the results of the DFT calculations.

\begin{acknowledgments}
The authors thank Y. Yanagi, M. Yatsushiro, H. Ikeda, Y. Nomura, and M. Suzuki for fruitful discussions.
This work was done under the Special Postdoctoral Researcher Program at RIKEN, and supported by the RIKEN TRIP initiative (RIKEN Quantum, Advanced General Intelligence for Science Program, Many-Body Electron Systems), JSPS KAKENHI Grants Numbers JP23K03288, JP23H00091, JP24K00581, JP23H04869, JP23K20827, and JST FOREST (JPMJFR2366), and the grants of Special Project (IMS program 23IMS1101), and OML Project (NINS program No, OML012301) by the National Institutes of Natural Sciences. 
 \end{acknowledgments}

\appendix

\section{Combined symmetry-adapted multipole basis}

The combined SAMBs in the graphene belonging to the identity $A_{1g}$ and symmetry breaking $A_{2u}$ irreps. up to 2nd-neighbor bond-cluster are summarized in the upper and lower panels of Table~\ref{tbl_graphene_combined_samb}, respectively.
The interpretation of each SAMB is given in Sec.~\ref{sec_graphene_E_field}.

\begin{table*}[h]
  \caption{\label{tbl_graphene_combined_samb}
  Combined SAMBs in monolayer graphene belonging to the identity $A_{1g}$ and symmetry breaking $A_{2u}$ irreps. up to 2nd-neighbor bond-cluster.
  The abbreviation $[\mathbb{X}\otimes\mathbb{Y}]_{} = (\mathbb{X}_{u}\otimes\mathbb{Y}_{u} + \mathbb{X}_{v}\otimes\mathbb{Y}_{v})/\sqrt{2}$ is used for $E_{1g}$, $E_{2g}$, $E_{1u}$, and $E_{2u}$ irreps.
  }
  \begin{ruledtabular}
  \begin{tabular}{cccccc|ccccc}
  $(\mathcal{H}$, Cluster) & $l$ & $\Gamma$ & $\gamma$ & Symbol & Definition & $l$ & $\Gamma$ & $\gamma$ & Symbol & Definition \\
  \hline
  $(\braket{s|s}, {\rm C})$ & 0 & $A_{1g}$ & $-$ & $\mathbb{Z}_{1}^{(A_{1g})}$ & $\mathbb{Q}_{0,s}^{\rm (a)}\otimes\mathbb{Q}_{0}^{\rm (C)}$ \\
  \hline
  $(\braket{p|p}, {\rm C})$ & 0 & $A_{1g}$ & $-$ & $\mathbb{Z}_{2}^{(A_{1g})}$ & $\mathbb{Q}_{0,p}^{\rm (a)}\otimes\mathbb{Q}_{0}^{\rm (C)}$ & 0 & $A_{1g}$ & $-$ & $\mathbb{Z}_{4}^{(A_{1g})}$ & $\mathbb{Q}_{0,(1,1)}^{\rm (a)}\otimes\mathbb{Q}_{0}^{\rm (C)}$\\
  & 2 & $A_{1g}$ & $-$ & $\mathbb{Z}_{3}^{(A_{1g})}$ & $\mathbb{Q}_{u}^{\rm (a)}\otimes\mathbb{Q}_{0}^{\rm (C)}$ & 2 & $A_{1g}$ & $-$ & $\mathbb{Z}_{5}^{(A_{1g})}$ & $\mathbb{Q}_{u,(1,-1)}^{\rm (a)}\otimes\mathbb{Q}_{0}^{\rm (C)}$\\
  \hline
  $(\braket{s|s}, {\rm B}_{1})$ & 0 & $A_{1g}$ & $-$ & $\mathbb{Z}_{6}^{(A_{1g})}$ & $\mathbb{Q}_{0,s}^{\rm (a)}\otimes\mathbb{Q}_{0}^{\rm (B_{1})}$ \\
  \hline
  $(\braket{s|p}, {\rm B}_{1})$ & 0 & $A_{1g}$ & $-$ & $\mathbb{Z}_{7}^{(A_{1g})}$ & $[\mathbb{T}_{1,E_{1u}}^{\rm (a)}\otimes\mathbb{T}_{1,E_{1u}}^{\rm (B_{1})}]$ & 0 & $A_{1g}$ & $-$ & $\mathbb{Z}_{18}^{(A_{1g})}$ & $[\mathbb{T}_{1,E_{1u},(1,0)}^{\rm (a)}\otimes\mathbb{T}_{1,E_{1u}}^{\rm (B_{1})}]$\\
   &  &  &  &  &  & 2 & $A_{1g}$ & $-$ & $\mathbb{Z}_{19}^{(A_{1g})}$ & $[\mathbb{M}_{2,E_{1u},(1,-1)}^{\rm (a)}\otimes\mathbb{T}_{1,E_{1u}}^{\rm (B_{1})}]$\\
   \hline
   $(\braket{p|p}, {\rm B}_{1})$ & 0 & $A_{1g}$ & $-$ & $\mathbb{Z}_{8}^{(A_{1g})}$ & $\mathbb{Q}_{0,p}^{\rm (a)}\otimes\mathbb{Q}_{0}^{\rm (B_{1})}$ & 0 & $A_{1g}$ & $-$ & $\mathbb{Z}_{20}^{(A_{1g})}$ & $\mathbb{Q}_{0,(1,1)}^{\rm (a)}\otimes\mathbb{Q}_{0}^{\rm (B_{1})}$\\
                                &   & $A_{1g}$ & $-$ & $\mathbb{Z}_{9}^{(A_{1g})}$ & $[\mathbb{Q}_{2,E_{2g}}^{\rm (a)}\otimes\mathbb{Q}_{2,E_{2g}}^{\rm (B_{1})}]$ &   & $A_{1g}$ & $-$ & $\mathbb{Z}_{21}^{(A_{1g})}$ & $[\mathbb{Q}_{2,E_{2g},(1,-1)}^{\rm (a)}\otimes\mathbb{Q}_{2,E_{2g}}^{\rm (B_{1})}]$\\
                                 & 2 & $A_{1g}$ & $-$ & $\mathbb{Z}_{10}^{(A_{1g})}$ & $\mathbb{Q}_{u}^{\rm (a)}\otimes\mathbb{Q}_{0}^{\rm (B_{1})}$ & 2 & $A_{1g}$ & $-$ & $\mathbb{Z}_{22}^{(A_{1g})}$ & $\mathbb{Q}_{u,(1,-1)}^{\rm (a)}\otimes\mathbb{Q}_{0}^{\rm (B_{1})}$\\
  \hline
  $(\braket{s|s}, {\rm B}_{2})$ & 0 & $A_{1g}$ & $-$ & $\mathbb{Z}_{11}^{(A_{1g})}$ & $\mathbb{Q}_{0,s}^{\rm (a)}\otimes\mathbb{Q}_{0}^{\rm (B_{2})}$ & 0 & $A_{1g}$ & $-$ & $\mathbb{Z}_{23}^{(A_{1g})}$ & $-\mathbb{M}_{z,s,(1,-1)}^{\rm (a)}\otimes\mathbb{M}_{z}^{\rm (B_{2})}$\\
  \hline
  $(\braket{s|p}, {\rm B}_{2})$ & 0 & $A_{1g}$ & $-$ & $\mathbb{Z}_{12}^{(A_{1g})}$ & $[\mathbb{Q}_{1,E_{1u}}^{\rm (a)}\otimes\mathbb{Q}_{1,E_{1u}}^{\rm (B_{2})}]$ & 0 & $A_{1g}$ & $-$ & $\mathbb{Z}_{24}^{(A_{1g})}$ & $[\mathbb{Q}_{1,E_{1u},(1,0)}^{\rm (a)}\otimes\mathbb{Q}_{1,E_{1u}}^{\rm (B_{2})}]$\\
                                &   & $A_{1g}$ & $-$ & $\mathbb{Z}_{13}^{(A_{1g})}$ & $[\mathbb{T}_{1,E_{1u}}^{\rm (a)}\otimes\mathbb{T}_{1,E_{1u}}^{\rm (B_{2})}]$ &   & $A_{1g}$ & $-$ & $\mathbb{Z}_{25}^{(A_{1g})}$ & $[\mathbb{T}_{1,E_{1u},(1,0)}^{\rm (a)}\otimes\mathbb{T}_{1,E_{1u}}^{\rm (B_{2})}]$\\
   &  &  &  &  &  & 2 & $A_{1g}$ & $-$ & $\mathbb{Z}_{26}^{(A_{1g})}$ & $[\mathbb{G}_{2,E_{1u},(1,-1)}^{\rm (a)}\otimes\mathbb{Q}_{1,E_{1u}}^{\rm (B_{1})}]$\\
   &  &  &  &  &  &   & $A_{1g}$ & $-$ & $\mathbb{Z}_{27}^{(A_{1g})}$ & $[\mathbb{M}_{2,E_{1u},(1,-1)}^{\rm (a)}\otimes\mathbb{T}_{1,E_{1u}}^{\rm (B_{1})}]$\\
  \hline
  $(\braket{p|p}, {\rm B}_{2})$ & 0 & $A_{1g}$ & $-$ & $\mathbb{Z}_{14}^{(A_{1g})}$ & $\mathbb{Q}_{0,p}^{\rm (a)}\otimes\mathbb{Q}_{0}^{\rm (B_{2})}$ & 0 & $A_{1g}$ & $-$ & $\mathbb{Z}_{28}^{(A_{1g})}$ & $\mathbb{Q}_{0,(1,1)}^{\rm (a)}\otimes\mathbb{Q}_{0}^{\rm (B_{2})}$\\
                                &   & $A_{1g}$ & $-$ & $\mathbb{Z}_{15}^{(A_{1g})}$ & $[\mathbb{Q}_{2,E_{2g}}^{\rm (a)}\otimes\mathbb{Q}_{2,E_{2g}}^{\rm (B_{2})}]$ &   & $A_{1g}$ & $-$ & $\mathbb{Z}_{29}^{(A_{1g})}$ & $[\mathbb{Q}_{2,E_{2g},(1,-1)}^{\rm (a)}\otimes\mathbb{Q}_{2,E_{2g}}^{\rm (B_{2})}]$\\
                                &   & $A_{1g}$ & $-$ & $\mathbb{Z}_{16}^{(A_{1g})}$ & $\mathbb{M}_{z}^{\rm (a)}\otimes\mathbb{M}_{z}^{\rm (B_{2})}$ &   & $A_{1g}$ & $-$ & $\mathbb{Z}_{30}^{(A_{1g})}$ & $[\mathbb{T}_{2,E_{2g},(1,0)}^{\rm (a)}\otimes\mathbb{T}_{2,E_{2g}}^{\rm (B_{2})}]$\\
                                & 2 & $A_{1g}$ & $-$ & $\mathbb{Z}_{17}^{(A_{1g})}$ & $\mathbb{Q}_{u}^{\rm (a)}\otimes\mathbb{Q}_{0}^{\rm (B_{2})}$ & & $A_{1g}$ & $-$ &  $\mathbb{Z}_{31}^{(A_{1g})}$ & $-\mathbb{M}_{z,p,(1,-1)}^{\rm (a)}\otimes\mathbb{M}_{z}^{\rm (B_{2})}$\\
  &  &  &  &  &  &   & $A_{1g}$ & $-$ & $\mathbb{Z}_{32}^{(A_{1g})}$ & $-\mathbb{M}_{z,a,(1,1)}^{\rm (a)}\otimes\mathbb{M}_{z}^{\rm (B_{2})}$\\
  &  &  &  &  &  & 2 & $A_{1g}$ & $-$ & $\mathbb{Z}_{33}^{(A_{1g})}$ & $\mathbb{Q}_{u,(1,-1)}^{\rm (a)}\otimes\mathbb{Q}_{0}^{\rm (B_{2})}$\\
  &  &  &  &  &  &   & $A_{1g}$ & $-$ & $\mathbb{Z}_{34}^{(A_{1g})}$ & $-[\mathbb{M}_{3,E_{2g},(1,-1)}^{\rm (a)}\otimes\mathbb{T}_{2,E_{2g}}^{\rm (B_{2})}]$\\
  &  &  &  &  &  &   & $A_{1g}$ & $-$ & $\mathbb{Z}_{35}^{(A_{1g})}$ & $\mathbb{M}_{3,A_{2g},(1,-1)}^{\rm (a)}\otimes\mathbb{M}_{z}^{\rm (B_{2})}$\\
  \hline\hline
  $(\mathcal{H}$, Cluster) & $l$ & $\Gamma$ & $\gamma$ & Symbol & Definition & $l$ & $\Gamma$ & $\gamma$ & Symbol & Definition \\
  \hline
  $(\braket{s|p}, {\rm C})$ & 1 & $A_{2u}$ & $-$ & $\mathbb{Z}_{1}^{(A_{2u})}$ & $\mathbb{Q}_{z}^{\rm (a)}\otimes\mathbb{Q}_{0}^{\rm (C)}$ & 1 & $A_{2u}$ & $-$ & $\mathbb{Z}_{2}^{(A_{2u})}$ & $\mathbb{Q}_{z, (1,0)}^{\rm (a)}\otimes\mathbb{Q}_{0}^{\rm (C)}$ \\
  \hline
  $(\braket{s|s}, {\rm B}_{1})$ &  &  & & & & $1$ & $A_{2u}$ & $-$ & $\mathbb{Z}_{8}^{(A_{2u})}$ & $[\mathbb{M}_{1,s,E_{1g}, (1,-1)}^{\rm (a)}\otimes\mathbb{T}_{1,E_{1u}}^{\rm (B_{1})}]$  \\
  \hline
  $(\braket{s|p}, {\rm B}_{1})$ & 1 & $A_{2u}$ & $-$ & $\mathbb{Z}_{3}^{(A_{2u})}$ & $\mathbb{Q}_{z}^{\rm (a)}\otimes\mathbb{Q}_{0}^{\rm (B_{1})}$ & 1 & $A_{2u}$ & $-$ & $\mathbb{Z}_{9}^{(A_{2u})}$ & $\mathbb{Q}_{z, (1,0)}^{\rm (a)}\otimes\mathbb{Q}_{0}^{\rm (B_{1})}$ \\
   &  &  & & & &  & $A_{2u}$ & $-$ & $\mathbb{Z}_{10}^{(A_{2u})}$ & $[\mathbb{G}_{2,E_{1u},(1,-1)}^{\rm (a)}\otimes\mathbb{Q}_{2,E_{2g}}^{\rm (B_{1})}]$ \\
   \hline
   $(\braket{p|p}, {\rm B}_{1})$ & $1$ & $A_{2u}$ & $-$ & $\mathbb{Z}_{4}^{(A_{2u})}$ & $[\mathbb{M}_{1,E_{1g}}^{\rm (a)}\otimes\mathbb{T}_{1,E_{1u}}^{\rm (B_{1})}]$ & 1 & $A_{2u}$ & $-$ & $\mathbb{Z}_{11}^{(A_{2u})}$ & $[\mathbb{M}_{1,p,E_{1g},(1,-1)}^{\rm (a)}\otimes\mathbb{T}_{1, E_{1u}}^{\rm (B_{1})}]$ \\
   &  &  & & & &  & $A_{2u}$ & $-$ & $\mathbb{Z}_{12}^{(A_{2u})}$ & $[\mathbb{M}_{1,a,E_{1g},(1,1)}^{\rm (a)}\otimes\mathbb{T}_{1, E_{1u}}^{\rm (B_{1})}]$ \\
    &  &  & & & &  & $A_{2u}$ & $-$ & $\mathbb{Z}_{13}^{(A_{2u})}$ & $\mathbb{M}_{3,B_{2g},(1,-1)}^{\rm (a)}\otimes\mathbb{T}_{3a}^{\rm (B_{1})}$ \\
      &  &  & & & &  & $A_{2u}$ & $-$ & $\mathbb{Z}_{14}^{(A_{2u})}$ & $[\mathbb{T}_{2,E_{1g},(1,0)}^{\rm (a)}\otimes\mathbb{T}_{1, E_{1u}}^{\rm (B_{1})}]$ \\
     &  &  & & & & 3 & $A_{2u}$ & $-$ & $\mathbb{Z}_{15}^{(A_{2u})}$ & $[\mathbb{M}_{3,E_{1g},(1,1)}^{\rm (a)}\otimes\mathbb{T}_{1, E_{1u}}^{\rm (B_{1})}]$ \\
  \hline
     $(\braket{s|s}, {\rm B}_{2})$ & & & & & & 1 & $A_{2u}$ & $-$ & $\mathbb{Z}_{16}^{(A_{2u})}$ & $[\mathbb{M}_{1,s,E_{1g}, (1,-1)}^{\rm (a)}\otimes\mathbb{T}_{1,E_{1u}}^{\rm (B_{2})}]$ \\
   \hline
    $(\braket{s|p}, {\rm B}_{2})$ & $1$ & $A_{2u}$ & $-$ & $\mathbb{Z}_{5}^{(A_{2u})}$ & $\mathbb{Q}_{z}^{\rm (a)}\otimes\mathbb{Q}_{0}^{\rm (B_{2})}$ & 1 & $A_{2u}$ & $-$ & $\mathbb{Z}_{17}^{(A_{2u})}$ & $\mathbb{Q}_{z, (1,0)}^{\rm (a)}\otimes\mathbb{Q}_{0}^{\rm (B_{2})}$\\
   &  &  & & & &  & $A_{2u}$ & $-$ & $\mathbb{Z}_{18}^{(A_{2u})}$ & $[\mathbb{G}_{2,E_{1u},(1,-1)}^{\rm (a)}\otimes\mathbb{Q}_{2,E_{2g}}^{\rm (B_{2})}]$ \\
    &  &  & & & &  & $A_{2u}$ & $-$ & $\mathbb{Z}_{19}^{(A_{2u})}$ & $\mathbb{M}_{0,(1,1)}^{\rm (a)}\otimes\mathbb{M}_{z}^{\rm (B_{2})}$ \\
    &  &  & & & &  & $A_{2u}$ & $-$ & $\mathbb{Z}_{20}^{(A_{2u})}$ & $[\mathbb{M}_{2,E_{1u},(1,-1)}^{\rm (a)}\otimes\mathbb{T}_{2,E_{2g}}^{\rm (B_{2})}]$ \\
    &  &  & & & & 3 & $A_{2u}$ & $-$ & $\mathbb{Z}_{21}^{(A_{2u})}$ & $\mathbb{M}_{u,(1,-1)}^{\rm (a)}\otimes\mathbb{M}_{z}^{\rm (B_{2})}$ \\
   \hline
     $(\braket{p|p}, {\rm B}_{2})$ & $1$ & $A_{2u}$ & $-$ & $\mathbb{Z}_{6}^{(A_{2u})}$ & $[\mathbb{Q}_{2,E_{1g}}^{\rm (a)}\otimes\mathbb{Q}_{1,E_{1u}}^{\rm (B_{2})}]$ & 1 & $A_{2u}$ & $-$ & $\mathbb{Z}_{22}^{(A_{2u})}$ & $[\mathbb{Q}_{2,E_{1g},(1,-1)}^{\rm (a)}\otimes\mathbb{Q}_{1,E_{1u}}^{\rm (B_{2})}]$ \\
   &  & $A_{2u}$ & $-$ & $\mathbb{Z}_{7}^{(A_{2u})}$ & $[\mathbb{M}_{1,E_{1g}}^{\rm (a)}\otimes\mathbb{T}_{1,E_{1u}}^{\rm (B_{2})}]$ &  & $A_{2u}$ & $-$ & $\mathbb{Z}_{23}^{(A_{2u})}$ & $[\mathbb{G}_{1,E_{1g}, (1,0)}^{\rm (a)}\otimes\mathbb{Q}_{1,E_{1u}}^{\rm (B_{2})}]$ \\
    &  &  & & & &  & $A_{2u}$ & $-$ & $\mathbb{Z}_{24}^{(A_{2u})}$ & $[\mathbb{M}_{1,p,E_{1g},(1,-1)}^{\rm (a)}\otimes\mathbb{T}_{1,E_{1u}}^{\rm (B_{2})}]$ \\
    &  &  & & & &  & $A_{2u}$ & $-$ & $\mathbb{Z}_{25}^{(A_{2u})}$ & $[\mathbb{M}_{1,a,E_{1g},(1,1)}^{\rm (a)}\otimes\mathbb{T}_{1, E_{1u}}^{\rm (B_{2})}]$ \\
    &  &  & & & &  & $A_{2u}$ & $-$ & $\mathbb{Z}_{26}^{(A_{2u})}$ & $\mathbb{M}_{3,B_{2g},(1,-1)}^{\rm (a)}\otimes\mathbb{T}_{3a}^{\rm (B_{2})}$ \\
    &  &  & & & &  & $A_{2u}$ & $-$ & $\mathbb{Z}_{27}^{(A_{2u})}$ & $[\mathbb{T}_{2,E_{1g},(1,0)}^{\rm (a)}\otimes\mathbb{T}_{1, E_{1u}}^{\rm (B_{2})}]$ \\
    &  &  & & & & 3 & $A_{2u}$ & $-$ & $\mathbb{Z}_{28}^{(A_{2u})}$ & $[\mathbb{M}_{3,E_{1g},(1,1)}^{\rm (a)}\otimes\mathbb{T}_{1, E_{1u}}^{\rm (B_{2})}]$ \\
  \end{tabular}
  \end{ruledtabular}
\end{table*}

\bibliographystyle{apsrev4-2}

\end{document}